\documentclass{article}
\usepackage{graphicx}  % 允许插入图片
\usepackage{amsmath, amssymb}  % 数学公式支持
\usepackage{hyperref}  % 超链接支持
\usepackage{booktabs}
\usepackage{subfigure}
\usepackage[authoryear]{natbib}
\title{Estimating Nodal Spreading Influence Using Partial Temporal Network}
\author{Tianrui Mao, Shilun Zhang, Alan Hanjalic, Huijuan Wang \\
    Delft University of University \\
    \texttt{T.Mao@tudelft.nl, H.Wang@tudelft}
}
\date{\today}

\begin{document}
\maketitle

\begin{abstract}
Temporal networks, whose links are activated or deactivated over time, are used to represent complex systems such as social interactions or collaborations occurring at specific times. Such networks facilitate the spread of information and epidemics. The average number of nodes infected via a spreading process on a network starting from a single seed node over a given period is called the influence of that node. In this paper, we address the question of how to utilize the partially observed temporal network (local and of short duration) around each node, to estimate the ranking of nodes in spreading influence on the full network over a long period. This is essential for target marketing and epidemic/misinformation mitigation where only partial network information is possibly accessible. This would also enable us to understand which network properties of a node, observed locally and shortly after the start of the spreading process, determine its influence. We systematically propose a set of nodal centrality metrics based on partial temporal network information, encoding diverse properties of (time-respecting) walks. It is found that distinct centrality metrics perform the best in estimating nodal influence depending on the infection probability of the spreading process. For a broad range of the infection probability, a node tends to be influential if it can reach many distinct nodes via time-respecting walks and if these nodes can be reached early in time. We find and explain why the proposed metrics generally outperform classic centrality metrics derived from both full and partial temporal networks.
\end{abstract}


\section{Introduction}\label{introduction}

Human social interactions or contacts usually occur at specific times instead of constantly and can be represented as temporal networks, where links are activated/deactivated over time. Spreading processes have been widely used to model the propagation of epidemics, the formation of opinion, and the cascade of failures on temporal networks. The average number of nodes that are infected via a spreading process on a network starting from a single seed node at $t=0$ over a given period $[1,\tau]$ is referred as the influence of this node. Recently, centrality metrics, or equivalently network properties of a node, have been proposed for temporal networks \citep{kim2012temporal,tsalouchidou2020temporal,ogura2017katz,praprotnik2015spectral}. Centrality metrics of nodes derived from the full temporal network underlying the spreading process observed within the same period $[1,\tau]$ are used to estimate the ranking of nodes in spreading influence \citep{bi2021temporal,wu2022temporal}. 

In practice, it is desirable to be able to estimate the ranking of nodes in influence over the long-term $[1,\tau]$, early in time, e.g., at $\phi\tau$ where $\phi \in (0,1)$, when the temporal network is only observable within a short-term $[1,\phi\tau]$. For instance, in the context of viral marketing, being able to identify the most influential users based on limited initial interactions allows companies to target their marketing efforts effectively, maximizing impact before the full spread of the campaign is visible. In this study, our objective is to utilize partially observed temporal network (local and early in time) around a node, to forecast the overall spreading influence of the node in the long term. This would help us to understand to what extent local properties of a node in the early time determine its spreading influence in the long term. It could also inspire the choice of the optimal seed node and the design of intervention strategies in the spreading process of information, epidemics, and opinion.  

In this work, we consider the discrete-time Susceptible-Infected (SI) spreading process \citep{hurley2006basic} on a temporal network. In the SI spreading process, each node can be in one of two possible states at any time: susceptible or infected. Initially at discrete time step $t=0$, a seed gets infected whereas all the other nodes are susceptible. At any time step $t$, a susceptible node could get infected by an infected node with an infection probability $\beta$ if the two nodes have an interaction or contact. If a node gets infected at time $t$, it could infect other susceptible nodes that it contacts since time $t+1$. The spreading influence of a node is defined as the average number of infected nodes (also known as the outbreak size) till time $t=\tau$ when the node is chosen as the only seed node of the SI epidemic spreading. We systematically explore how partial temporal network information around a node may contribute to the prediction of the ranking of nodes in spreading influence. Specifically, we address the following generalized nodal influence prediction problem: given the partial temporal network $\mathcal{G}_{i}(\phi,m)$ of each target node $i$ that is the temporal network $G$ observed in the early period $[1,\phi\tau]$ and within $m$ hops from $i$, how to estimate the ranking of nodes in spreading influence on the whole temporal network $G$ over the longer period $[1,\tau]$? The partial temporal network $\mathcal{G}_i(\phi,m)$ contains the set of nodes $\mathcal{V}_i(m)$ that includes node $i$ and nodes that are within $m$ hops\footnote{The hopcount between two nodes in a network is the number of links contained in the shortest path between the two nodes.} from $i$ in the unweighted aggregated network of the temporal network $\mathcal{G}$ observed within $[1,\phi\tau]$ and all contacts of the temporal network $G$ that occur within $[1,\phi\tau]$ and among $\mathcal{V}_i(m)$.
%Existing nodal centrality metrics in temporal networks characterize nodal centrality based on a certain nodal network property. However, they do not fully consider how the selected nodal network property is related to the spreading process originating from the node. 

To solve this problem, we design three nodal centrality metrics derived from the partial temporal network $\mathcal{G}_{i}(\phi,m)$ to predict the influence of the node. Centrality metrics are defined to quantify various network properties of a node, respectively. Centrality metrics have already been proposed to capture the properties of a node in relation to the shortest paths in a static network or the shortest time-respecting paths in a temporal network. For example, temporal closeness quantifies how close a node is connected to other nodes via time-respecting paths. Such metrics could be limited in estimating nodal spreading influence, because the spread of information from a seed node to any other node is not necessarily through the shortest time-respecting path, but possibly through any time-respecting path. This motivates us to define centrality metrics that 
systematically capture how well the partial temporal network around a node is connected via time-respecting walks and via walks in the corresponding aggregated network $\mathcal{G}^w_i(\phi,m)$, thus accounting all possible spreading trajectories starting from the seed node. Using centrality metrics based on partial network information to predict nodal influence also enables the exploration of how large $m$, thus to what extent relatively local information, is actually needed to provide a desirable prediction performance. Earlier research \citep{meghanathan2017computationally,zhang2024predicting,zhang2018link} have demonstrated in static networks that a centrality metric derived from full network, such as betweenness, which involves a high computational complexity, exhibits a significant correlation with centrality metrics obtained from local neighborhoods. Therefore, predicting nodal influence might not necessarily require extensive macroscopic neighborhood information.

It is found that the proposed centrality metrics derived from partial network information, in general, outperform classic centrality metrics utilizing either full or partial temporal network information. We further explain the performance of these methods as well as their dependency on the infection probability of the spreading process. For a broad range of the infection probability, a node tends to be influential if it could reach many distinct nodes via time-respecting walks and if these nodes could be reached early in time.

The remainder of this paper is organized as follows. In Section \ref{Temporal network representation}, (partial) temporal networks and their
weighted aggregated networks are defined. In Section \ref{local centrality metric}, we introduce our walk-based nodal centrality metrics as well as classic centrality metrics. In Section \ref{Evaluation of prediction quality}, methods including datasets used to evaluate the prediction quality of aforementioned metrics are explained. In Section \ref{Performance analysis}, we analyze and explain the performance of  
 those metrics in multiple real-world networks and their randomized networks. Section \ref{Conclusions and future work} summarizes our findings and discusses possible future works.

\section{Temporal network representation}\label{Temporal network representation}
A temporal network measured within an observation window $[1,\tau]$ at discrete times can be represented as a sequence of network snapshots $G=\{G_1,G_2,...,G_{\tau}\}$. The snapshot $G_t=(V;E_t)$ at time step $t$ has $V$ and $E_t$ being the set of nodes and contacts, respectively. The number of nodes in $V$ is represented as $N$. If node $i$ and $j$ have an interaction or a contact at time step $t$, $(i,j)\in E_t$. Here, we assume all snapshots share the same set of nodes, i.e., $V$. A temporal network can also be described by a three-dimensional binary adjacency matrix $A_{N \times N \times \tau}$, where each element $a_{i,j,t} = 1$ if there is a contact between node i and j at time step t, or else $a_{i,j,t}$ = 0. 

A weighted aggregated network $G^w$ can be derived from a temporal network $G$ by aggregating contacts over time window $[1,\tau]$. The links in the time aggregated network $G^w$ are defined as $E=\cup_{t=1}^{\tau}E_t$. That is, a pair of nodes is connected with a link in $G^w$ if at least one contact occurs between them in the temporal network. Each link $(i,j)$ in $G^w$ is associated with a weight $w_{i,j}$ counting the total number of contacts between node $i$ and $j$ in $G$. The weighted aggregated network $G^w$ can therefore be described by a weighted adjacency matrix $W_{N \times N}$, with its element $w_{i,j} = \sum_{t=1}^{\tau} a_{i,j,t}$. 

\begin{figure*}[htbp]
  \centering
  \includegraphics[width=1\textwidth]{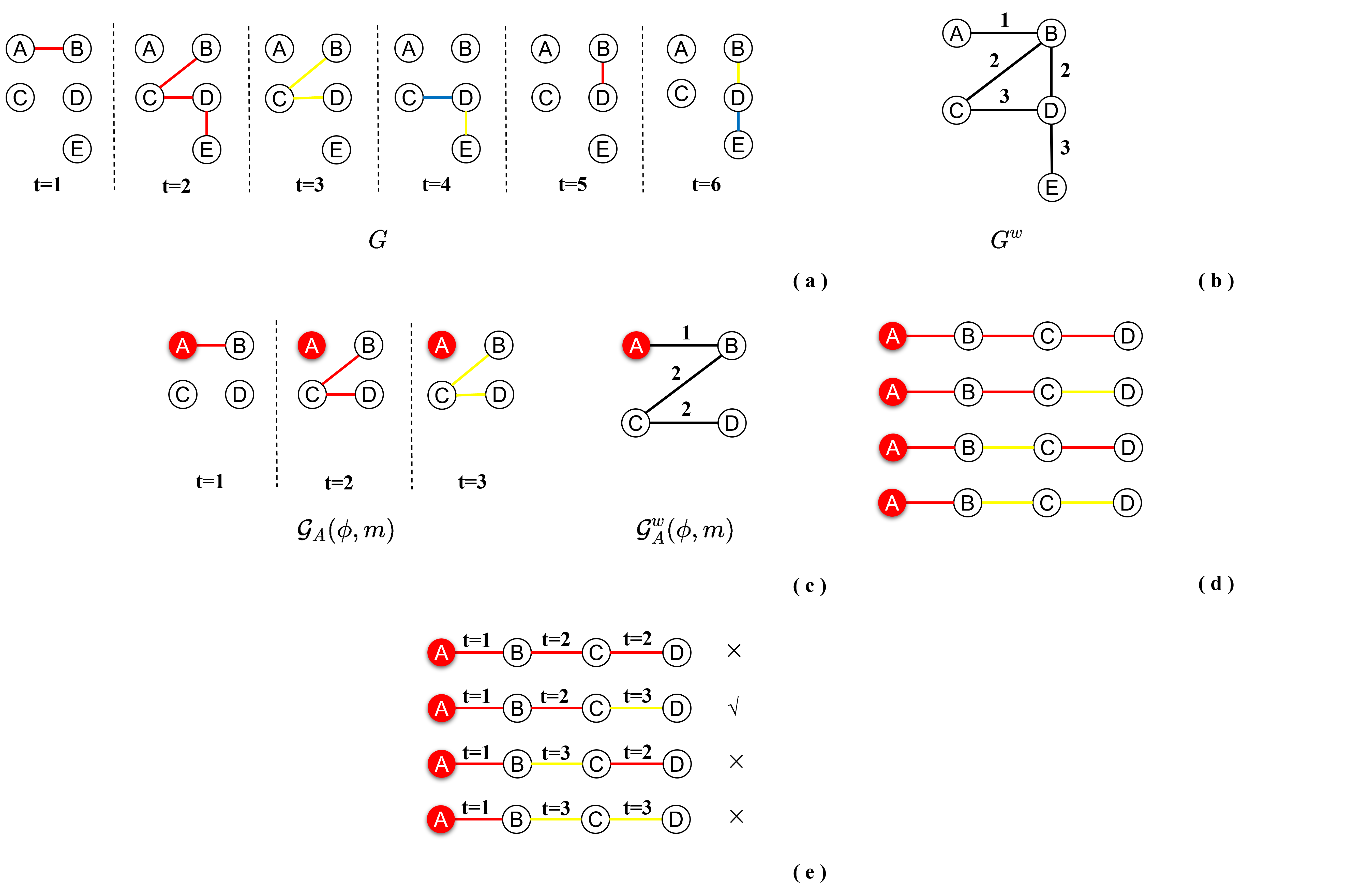} % 图片文件名为 example-image，可以替换为你的图片文件名
  \caption{(a) A temporal network $G$ with 5 nodes and 6 time steps. The first, second, and third contacts between the same pair of nodes are marked in red, yellow, and blue, respectively. (b) The aggregated network $G^w$ of $G$ along with its link weight. (c) The partial temporal network $\mathcal{G}_{A}(\phi,m)$ where $\phi$ = 0.5 and $m$ = 3, observed around node $A$ and its corresponding weighted aggregated network $\mathcal{G}_{A}^w(\phi,m)$. (d) The list of all 3-hop walks between node $A$ and $D$ in $\mathcal{G}_{A}^w(\phi,m)$. (e) For each walk listed in (d), time stamp of each link is added and the walk is marked as $\times$ if it is not a time-respecting walk and as $\checkmark$ if it is time-respecting.}
  \label{scheme}
\end{figure*}

The partial temporal network $\mathcal{G}_{i}(\phi,m)$ observed around each target node $i$ in the early period $[1,\phi\tau]$ and within $m$ hops from $i$ will be used to estimate the ranking of nodes in spreading influence on the whole temporal network $G$ over the longer period $[1,\tau]$. Figure \ref{scheme} (a)-(c) shows the example of a temporal network $G$, its weighted aggregated network $G^w$, the partial temporal network $\mathcal{G}_{A}(\phi,m)$ observed around node $A$ and the corresponding weighted aggregated network $\mathcal{G}_{A}^w(\phi,m)$, where $\phi$ = 0.5 and $m$ = 3. 

\section{Centrality metrics}\label{local centrality metric}

We first design three walk-based nodal centrality metrics, namely weighted degree mass, (time-scaled) temporal degree mass, and (time-scaled) temporal reachability, to capture properties of the partial temporal network observed around a node. Each metric is firstly defined based on a full temporal network $G$ or its aggregated network $G^w$ and then adapted for partial temporal network. Each metric derived from the partial network is used to estimate the rank of the nodes in influence. In order to evaluate the performance of the proposed walk-based metrics, we also introduce a set of classic centrality metrics and explain how these metrics could be derived from partial or full temporal networks or their aggregated networks, respectively, to estimate nodal influence.  

\subsection{Weighted degree mass}
We firstly propose the definition of the mth-order weighted degree mass $d_i^{(m)}$ of a node $i$ in the weighted aggregated network $G^w$ of a temporal network $G$ with its weighted adjacency matrix $W$ as 
\begin{equation}
d_i^{(m)}=\sum_{k=1}^{m}{(W^ku)_i} \label{1}
\end{equation}
where $u=(1,1,...,1)^T$ is the all-one vector. Each element $W^k_{i,j}$ represents the total number of distinct k-hop ($k\leq m$) walks\footnote{A k-hop walk between node $n_0$ and $n_k$ in a weighted aggregated network is a succession of links $(n_0,n_1),(n_1,n_2),..(n_k-1,n_k)$.} between node $i$ and $j$ ($i$, $j$ can be the same node) in $G^w$ when interpreting the weight $w_{i,j}$ of each node pair as the number of links between the node pair. Therefore, $(W^ku)_i$ counts the total number of distinct k-hop walks starting from node $i$. The mth-order degree mass $d_i^{(m)}$ of a node $i$ represents the total number of walks within m-hops starting from the node $i$. This definition generalizes the original degree mass definition that has been proposed for unweighted networks by replacing the unweighted adjacency matrix with a weighted one $W$ \citep{li2015correlation}. 

In our context, only the partial temporal network $\mathcal{G}_{i}(\phi,m)$ observed around node $i$ within a short period $[1,\phi\tau]$ is known, and its aggregated network is $\mathcal{G}^w_{i}(\phi,m)$ with adjacency matrix $\mathcal{W}_{i}(\phi,m)$. We propose to consider the mth-order weighted degree mass $d_i^{(\phi, m)}$ in the aggregated partial network $\mathcal{G}^w_{i}(\phi,m)$, i.e., 

\begin{equation}
d_i^{(\phi,m)}=\sum_{k=1}^{m}{(\mathcal{W}^k_{i}(\phi,m)u)_i} \label{2}
\end{equation}
It represents the total number of distinct walks within $m$-hop starting from node $i$ in the aggregated partial network observed around $i$ within $m$ hops during $[1,\phi\tau]$. An example of $m$-hop walks in $\mathcal{G}^w_{i}(\phi,m)$ between two nodes is shown in figure \ref{scheme}(d). Correspondingly, we propose to use the $m$th-order weighted degree mass $d_i^{(\phi, m)}$ to estimate the influence of the node. For any link in $G^w$, its weight is larger or equal to its weight in $\mathcal{G}^w_{i}(\phi,m)$. Hence, $d_i^{(\phi,m)} \leq d_i^{(m)}$.

\subsection{Temporal degree mass}
Furthermore, we propose the mth-order temporal degree mass $\delta_i^{(m)}$ of a node $i$ in a temporal network $G$ as the total number of time-respecting walks\footnote{A k-hop time-respecting walk from $n_0$ at $t=0$ to $n_k$ in a temporal network is a succession of contacts $(n_0,n_1,t_1),(n_1,n_2,t_2),..(n_k-1,n_k,t_k)$ that follow the time order $0<t_1<t_2,...,<t_k$.} within $m$-hops starting from node $i$ at time $t=1$. An example of $m$-hop time-respecting walks in the partial temporal network $\mathcal{G}_{i}(\phi,m)$ between two nodes is shown in figure \ref{scheme} (e). The total number of time-respecting walks within $m$-hops starting from node $i$ at time $t=1$ to any node $j$ is upper-bounded by the total number of walks within $m$-hops from $i$ to $j$ in the weighted aggregated network $\Psi^{(m)}_{i,j}$ where $\Psi^{(m)}=\sum_{k=1}^{m}{W^k}$, as exemplified in figure \ref{scheme} (d) and (e). We use vector $B_{i,j}^{(m)}$, a row vector of length $\Psi^{(m)}_{i,j}$ with binary elements to indicate whether each of the $\Psi^{(m)}_{i,j}$ walks in the weighted aggregated network is time-respecting or not when the time information of each link in the walk is taken into account. Hence, the $m$th-order temporal degree mass $\delta_i^{(m)}$ of a node $i$ in a temporal network follows

\begin{equation}
\delta_i^{(m)}=\sum_{j=1}^{N}{\sum_{s=1}^{\Psi^{(m)}_{i,j}}B_{i,j}^{(m)}(s)}
%\delta_i^{(m)}=\sum_{k=1}^{m}{(\prod_{s=1}^{k}{A(s)}u)_{i}}
\end{equation}

A node $i$ with a high temporal degree mass $\delta_i^{(m)}$, thus well connected via many time-respecting walks within $m$-hops to other nodes may contribute to the high influence of the node. Beyond, if these walks have a short time duration, i.e., the destination nodes of these walks are reached at an early time, node $i$ tends to be more influential. This is because of the following. Only those nodes that can be reached by time-respective walks are possibly infected in the stochastic SI spreading process starting staring from seed $i$ at time $t=0$. If a node $j$ is infected, the spreading trajectory, i.e.,  the sequence of contacts via which j gets infected is a time-respecting path. Any time-respecting walk from $i$ to $j$, that is also a time-respecting path, is a possible epidemic spreading trajectory. Hence, a large number of time-respecting walks within $m$-hops from $i$ to $j$ suggests a high probability for $j$ to get infected in the spreading process starting from $i$. If $j$ gets infected earlier, more contacts that occur after its infection could spread the epidemic/information further from $j$. Hence, if a node $i$ has a high temporal degree mass $\delta_i^{(m)}$, thus likely a large number of time-respecting paths within $m$-hops to other nodes, and if these walks tend to have a short time duration such that other nodes may get infected further, it tends to have a high influence. Therefore, we propose to use the time-scaled temporal degree mass $\Delta_i^{(m)}$ of a node $i$ to estimate the influence of the node. The time-scaled temporal degree mass $\Delta_i^{(m)}$ is defined as 
\begin{equation}
\Delta_i^{(m)}=\sum_{j=1}^{N}{\sum_{s=1}^{\Psi^{(m)}_{i,j}}{B_{i,j}^{(m)}(s)\alpha^{\phi_{i,j}^{(m)}(s)}}}
\end{equation}

where the vector $\phi_{i,j}^{(m)}$ records the time duration\footnote{The time duration of a time-respecting walk equals the time of the last contact in the walk minus the starting time of epidemic spreading, which is 0 in our context.} of each walk from $i$ to $j$ within $m$-hops identified in the weighted aggregated network equipped with the time stamp of each link in the walk in case the walk is time-respecting and is infinity in case the walk is not time-respecting. The vector $\phi_{i,j}^{(m)}$ thus has the same length $\Psi^{(m)}_{i,j}$ as vector $B_{i,j}^{(m)}$, which equals the total number of walks from i to j within $m$-hops in the weighted aggregated network. Both vectors index the $\Psi^{(m)}_{i,j}$ walks in the same way. If the $s$-th walk in the aggregated network is time-respecting in the temporal network, i.e., $B_{i,j}^{(m)}{(s)}=1$, the contribution of this walk to the time-scaled temporal degree mass $\Delta_i^{(m)}$ is $\alpha^{\phi_{i,j}^{(m)}{(s)}}$, where $\alpha\in (0,1]$. When $\alpha<1$, a walk with a longer duration contributes less to the time-scaled temporal degree mass. When $\alpha=1$, the time-scaled temporal degree mass $\Delta_i^{(m)}=\delta_i^{(m)}$ equals the temporal degree mass, counting simply the total number of time-respecting walks within $m$-hops from $i$. 

In our context, we use the time-scaled temporal degree mass $\Delta_i^{(\phi, m)}$ derived from the partial temporal network $\mathcal{G}_{i}(\phi,m)$ to estimate the influence of the node $i$. %In this case, $\Psi^{(m)}_{i,j}$, $B_{i,j}^{(m)}$ and $\phi_{i,j}^{(m)$ are derived from the (aggregated) partial temporal network.

\subsection{Temporal reachability}
Besides time-respecting walks, the concept of the reachable node can also be utilized for designing centrality metrics. The temporal reachability $z_{i}^{(m)}$ of a node $i$ within $m$-hops in a temporal network $G$ is the number of distinct nodes that could be reached via time-respecting walks starting from node $i$ at $t=0$ within $m$-hops. Mathematically, 
\begin{equation}
z_{i}^{(m)}=\sum_{j=1}^{N}{\textbf{1}_{\sum_{s=1}^{\Psi^{(m)}_{i,j}}B_{i,j}^{(m)}(s)>0}}
\end{equation}
where the condition $\sum_{s=1}^{\Psi^{(m)}_{i,j}}B_{i,j}^{(m)}(s)>0$ is true if there is at least one time-respecting walk from $i$ to $j$ and the indicator function $\textbf{1}_{x}$ equals one if the condition $x$ is true or zero otherwise.

When the infection probability per contact is high, any time-respecting walk starting from node $i$ to $j$ within $m$-hops could lead to the infection of $j$ with a high probability. In this case, the temporal reachability (the number of nodes can be reached via walks) instead of the temporal degree mass (the number of distinct walks) could be more relevant in estimating nodal influence. Similarly, if a node $i$ has high temporal reachability $z_{i}^{(m)}$ and if each of these $z_{i}^{(m)}$ nodes has an earlier reached time\footnote{The reached time equals the duration when the node is firstly reached by the seed minus the starting time of epidemic spreading, which is 0 in our context.}, node $i$ tends to be influential. Therefore, we propose to use the time-scaled temporal reachability $Z_{i}^{(m)}$ of a node $i$ to estimate the influence of the node, if the global network $G$ is known. The time-scaled temporal reachability $Z_{i}^{(m)}$ is defined as 
\begin{equation}
Z_{i}^{(m)}=\sum_{j=1}^{N}{\textbf{1}_{\sum_{s=1}^{\Psi^{(m)}_{i,j}}B_{i,j}^{(m)}(s)>0} \alpha^{\min_{s}\phi_{i,j}^{(m)}(s)}}
\end{equation}

where the contribution of each reachable node $j$ is scaled by $\alpha^{\min_{s}\phi_{i,j}^{(m)}(s)}$, depending on the shortest time $\min_{s}\phi_{i,j}^{(m)}(s)$ that $j$ is reached. 
In our context, we use the time-scaled temporal reachability $Z_{i}^{(\phi,m)}$ derived from the partial temporal network $\mathcal{G}_{i}(\phi,m)$ to estimate the influence of the node $i$. The time-scaled temporal reachability $Z_{i}^{(\phi,m)}$ takes into account how many nodes are reachable via time-respective walks/paths in the partial temporal network as well as when each node is reached via the fastest time-respective path. 

These three proposed centrality metrics, derived from the full (aggregated) temporal network, respectively, follow

\begin{equation}
z_{i}^{(m)} \leq \delta_i^{(m)} \leq d_i^{(m)} \label{eq9}
\end{equation}

This is because not all the $d_i^{(m)}$ walks identified from the aggregated network are necessarily time-respecting and not each of the $\delta_i^{(m)}$ time-respecting walks reaches a unique destination node. Similarly,

\begin{equation}
Z_{i}^{(\phi,m)} \leq \Delta_i^{(\phi,m)} \leq d_i^{(\phi,m)} \label{eq10}
\end{equation}
when the same scaling $\alpha$ is considered.
 
\subsection{Classic centrality metrics}\label{Centrality metrics based on full temporal network}
Centrality metrics have been proposed in static networks and recently in temporal networks. We are interested in how the proposed centrality metrics, leveraging partial network information, perform in comparison with classic centrality metrics in influence estimation. We will firstly introduce 4 centrality metrics defined for the static networks. Each metric, derived from the unweighted aggregated network of temporal network $G$, or of partial temporal network $\mathcal{G}_{i}(\phi,m)$, as well as the average of this metric computed across all snapshots of $G$ or of $\mathcal{G}_{i}(\phi,m)$, will be used to predict nodal influence, respectively.

\begin{enumerate}
  \item[*]The betweenness centrality $b_i$ \citep{wang2008betweenness} of a node $i$ is the number of shortest paths between all pairs of nodes in the network that pass through the node $i$, 

  \begin{equation}
    b_i = \sum_{s \neq i \neq d \in V} \frac{\sigma_{sd}(i)}{\sigma_{sd}} 
  \end{equation}
  
  where $\sigma_{sd}(i)$ is the number of shortest paths that pass through node $i$ between node $s$ and node $d$, and $\sigma_{sd}$ is the total number of shortest paths between $s$ and $d$. Assuming that a unit packet is transmitted between
  each node pair via the shortest path, the betweenness $b_i$ is the total number of packets passing through node $i$. 
  
  \item[*]The closeness centrality $c_i$ \citep{sabidussi1966centrality} of a node $i$ measures how close a node is connected to all the others via the shortest path. It is commonly defined as   
  \begin{equation}
      c_i = \sum_{j \in V \backslash \{i\}}\frac{1}{{H_{i,j}}}
  \end{equation}
  where $H_{i,j}$ is the hopcount of the shortest path between nodes $i$ and $j$. 
  
  \item[*]The eigenvector centrality $x_i$ \citep{bonacich1972factoring} of node $i$ is the component of the principal eigenvector $x$ corresponding to node i and the principal eigenvector is the eigenvector corresponds to the largest eigenvalue $\lambda_1$ of the adjacency matrix $A$ of the static network. Hence, $x \lambda_1=Ax$. The eigenvector centrality $x_i$ of a node tends to be large if it has many neighbors and each neighbor has a large eigenvector centrality.
  
  \item[*]The PageRank centrality $P_i$ \citep{gleich2015pagerank} of node $i$ is the probability that node $i$ is visited by a random walker:
  \begin{equation}
      P_i = \frac{1 - \gamma}{N} + \gamma\sum_{j \in V \backslash \{i\}} \frac{A_{i,j}P_j}{k_{j}}
  \end{equation}
  where $\gamma$ is the probability for a walker to move to a random neighbor of the current node being visited, $1-\gamma$ is thus the probability for the walker to move to a random node and $k_{j}$ is the degree of node $j$. The parameter $\gamma$ is set to 0.85, which is a common choice for calculating the PageRank centrality.
\end{enumerate}
Furthermore, we introduce the temporal closeness centrality defined for temporal networks. It will be derived from the full temporal network $G$ and partial temporal network $\mathcal{G}_{i}(\phi,m)$ respectively to predict nodal influence.
\begin{enumerate}
  \item[*]Temporal closeness centrality $TC_i$ \citep{kim2012temporal} is defined analogously as closeness centrality. It measures how close node $i$ is connected to the other nodes via time-respecting paths. Specifically, it is defined as

  \begin{equation}
      TC_i = \sum_{j \in V \backslash \{i\}}\frac{1}{{TH_{i,j}}}
  \end{equation}
  where $TH_{i,j}$ is the hopcount of the shortest time-respecting path\footnote{The shortest time-respecting path is the time-respecting path with the minimum number of hopcount, or equivalently with the minimum number of contacts.} from node $i$ to $j$, as introduced in \citep{wu2016efficient}.
\end{enumerate}

\section{Evaluation of prediction quality}\label{Evaluation of prediction quality}
In this Section, we introduce the method to evaluate the quality of an influence prediction algorithm/metric. This entails the real-world networks to be used, the parameter choice of the spreading process and the partial temporal network, and the measures that quantify the prediction quality.
\subsection{Empirical networks}
The following real-world temporal networks will be considered to evaluate the aforementioned nodal spreading influence estimation methods.
\begin{enumerate}
  \item[*] HighSchool11\&12 \citep{fournet2014contact} record the physical contacts between students in a high school in Marseilles, France. These two datasets incorporate two different groups of students.
  \item[*] WorkPlace13\&15  \citep{genois2015data} capture the physical contacts between individuals in an office building in France. These two datasets originate from distinct sets of individuals.
  \item[*] Hyper-text\&SFHH \citep{isella2011s,cattuto2010dynamics} record the physical contacts among scientists during the 2009 conference of ACM Hypertext and SFHH. 
  \item[*] Sms\&Calls \citep{sapiezynski2019interaction} are obtained from the contacts via short messages and calls of the same set of mobile phones on campus, respectively. 
  \item[*] Manufacturing Emails \citep{kunegis2013konect} represents the internal email communication network between employees of a mid-sized manufacturing company.
\end{enumerate}

These networks record virtual or face-to-face contacts in the context of workplace, highschool, university, and academic conference at discrete time steps. The time steps at which there is no contact in the whole network have been deleted. This pre-processing has also been used in e.g. \citep{zou2022memory}. In this way, we focus solely on those time steps relevant to information diffusion, and exclude periods without any contact, possibly resulted from the inactive period during evenings or technical errors when measuring the network. Meanwhile, we observe that the aggregated networks of Sms and Calls are not fully connected. Hence, we extract the largest connected components of these two aggregated networks respectively, and consider only nodes within each largest connected component and contacts between these nodes. Basic statistics of selected empirical temporal networks after pre-processing are shown in table \ref{tab:real_world_networks}. The link density $p$ is the number of links in the aggregated network of a temporal network normalized by $N(N-1)/2$, i.e., the maximum possible number of links among the same set of $N$ nodes.

\begin{table}[h]
\centering
\caption{Basic statistics of each real-world temporal network considered: the number of nodes (N), the total number of contacts ($L$), the total number of timesteps (T), the type of contacts recorded, the link density ($p$), and the average of modularity $\Gamma$ of the largest connected component of the weighted network $G^w$ over all considered observation periods.}
\label{tab:real_world_networks}
\resizebox{\textwidth}{!}{% 调整表格大小为文本宽度
\begin{tabular}{cccccccc}
\toprule
 & N & L & T & Type & $p$ & $\Gamma$\\
\midrule
Workplace13 & 92 & 9827 & 7104 & Physical  & 0.180 & 0.592\\
Workplace15 & 217 & 78249 & 18488 & Physical  & 0.182 & 0.641\\
Highschool11 & 126 & 28561 & 5609  & Physical   & 0.217& 0.667\\
Highschool12 & 180 & 45047 & 11273  & Physical  & 0.138 & 0.754\\
Hyper-text & 113 & 20818 & 5246   & Physical   &  0.347  & 0.441\\
SFHH & 403 & 70261 & 3509  &  Physical  & 0.118 &  0.536\\
Sms (filtered) & 457 & 22152 & 21898 & Virtual  & 0.006 & 0.912\\
Calls (filtered) & 347 & 2676 & 2671 & Virtual  & 0.008 & 0.878\\
Manufacturing Emails & 167 & 82281 & 57791 & Virtual  & 0.234& 0.401\\
\bottomrule
\end{tabular}
}
\end{table}

\subsection{Experimental settings}\label{Experimental settings}

\begin{figure}[htbp]
  \centering
  \includegraphics[width=0.5\textwidth]{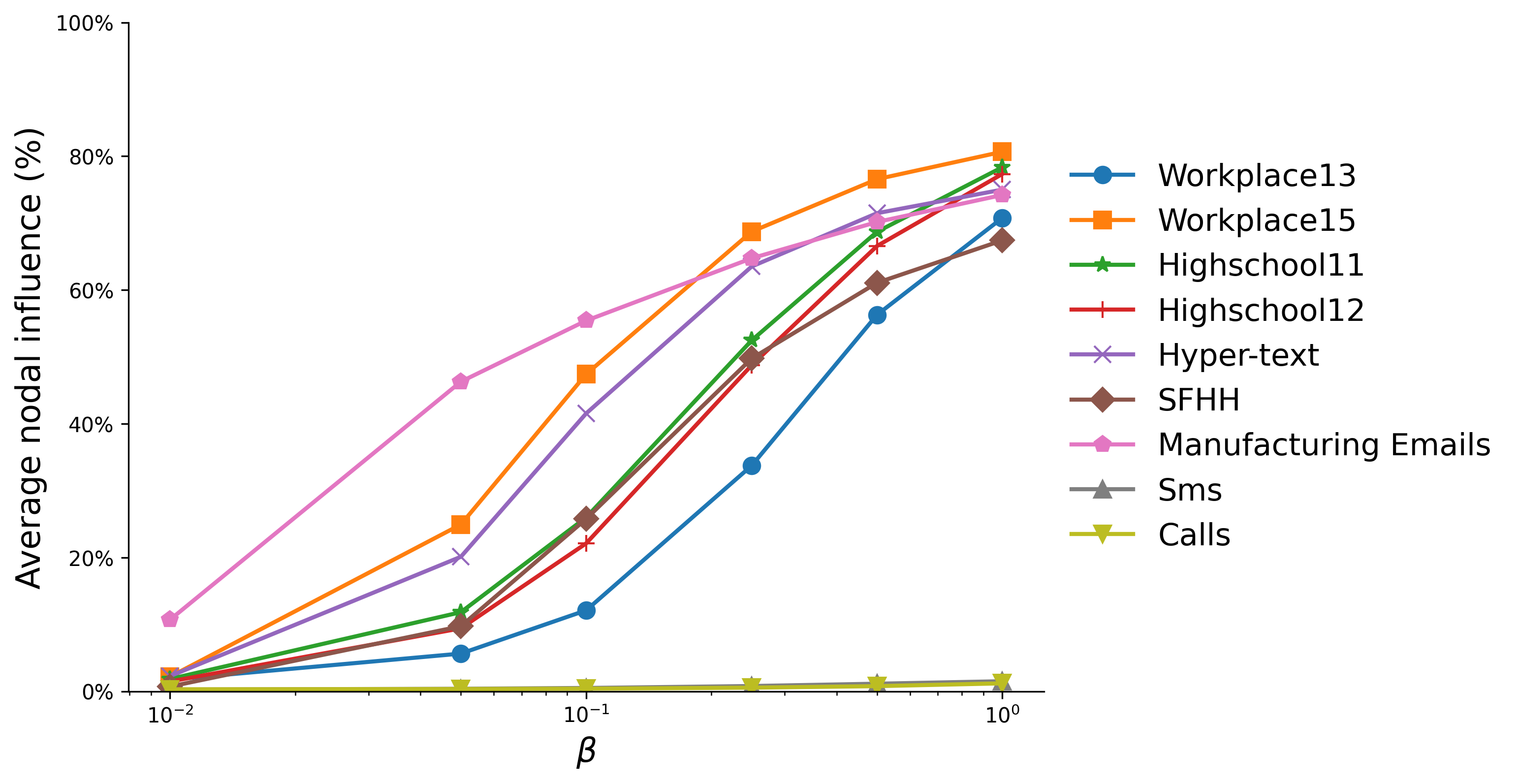} % 图片文件名为 example-image，可以替换为你的图片文件名
  \caption{Average nodal influence in each real-world network as a function of the infection probability $\beta$.}
  \label{Average nodal influence in each real-world network}
\end{figure}

Without loosing generality, we evaluate the performance of using the proposed nodal centrality metrics for influence prediction when the infection probability $\beta$ of the SI spreading is systematically examined across a broad range, i.e., $\beta \in \{0.01, 0.05, 0.1, 0.25, 0.5, 1\}$. When $\beta$ \textless\ 1, the SI process is stochastic, and the actual spreading influence of a node is derived as the average outbreak size over $500$ independent realizations of the SI process starting from this node. When we define the spreading process and propose nodal centrality metrics in Section \ref{introduction} and \ref{local centrality metric}, the starting time $t_0$ of the spreading is assumed to be $0$. We consider the following more general case. For each dataset in Table \ref{tab:real_world_networks}, we consider a set of possible starting times, i.e., $t_0\in\{0, T/8, T/4, 3T/8, T/2, 5T/8, 3T/4\}$. Given a starting time and a seed node, the influence of this node, i.e., the average outbreak size at $t_0 + \tau$ is considered, where $\tau=T/4$. Hence, the temporal network within $[t0, t0 + \tau]$ decides the influence of each node. The ranking of nodal influence is estimated via each of the centrality metrics we proposed based on the partial temporal network observed within $[t0, t0 + \phi\tau]$ and within $m$ hops around each node, where $\phi$ is chosen as $0.25$ and $0.5$, respectively and $m \in [1,2,3]$. For each proposed centrality metric, $\alpha \in (0,1]$ could be tuned to achieved the best performance. The average influence of a node over all seed nodes and possible starting times as a function of the infection probability $\beta$ in each real-world network is shown in figure \ref{Average nodal influence in each real-world network}. Network Sms and Call that have the lowest link density among all networks as shown in table \ref{tab:real_world_networks}. Correspondingly, they have the lowest average nodal influence, thus prevalence in the spreading process. 

\subsection{Prediction quality}
We introduce two measures to evaluate the quality of each proposed centrality metric in predicting the ranking of nodal influence. The actually influence of each node is represented by a vector $s$, whereas $\hat{s}$ records a given centrality metric for each node, representing the predicted influence.

Kendall’s correlation coefficient $Q_{k}{(\hat{s},s)}$ \citep{kendall1945treatment} measures the similarity between the ranking of nodes based on the predicted nodal influence $\hat{s}$, and the ranking of nodes based on the actual nodal influence vector $s$ obtained by SI simulation. A value of $1$ for $Q_{k}{(\hat{s},s)}$ indicates that the centrality metric gives the same node ranking as the ground truth nodal influence, while a value of $-1$ indicates that the two rankings are reversed. Kendall’s correlation coefficient is defined as:
  \begin{equation}
    Q_{k}{(\hat{s},s)} = \frac{n_c - n_d}{\sqrt{(n_c + n_d + O) \cdot (n_c + n_d + U)}}
  \end{equation}
  where $n_c$ and $n_d$ are the total numbers of node pairs that are concordant and discordant respectively, based on the influence $s$ and the predicted influence $\hat{s}$. For example, node pair $(i, j)$ is concordant if $(\hat{s}_i - \hat{s}_j)(s_i - s_j) > 0$, and is discordant if $(\hat{s}_i - \hat{s}_j)(s_i - s_j) < 0$. The number of node pairs that have the same actual influence but different predicted influence, i.e., $s_i = s_j$,$\hat{s}_i \neq \hat{s}_j$ is denoted by $O$ and $U$ is the number of node pairs that have the same predicted influence but different actual influence, i.e., $\hat{s}_i = \hat{s}_j$,$s_i \neq s_j$.

Recognition rate $Q_{r}{(\hat{s},s, f)}$ of top-$f\%$ measures the performance of a centrality metric in identifying the most influential $f\%$ nodes. It is calculated as the proportion of nodes that are present in both $V_f$, the set of top $f\%$ of nodes ranked by the predicted nodal influence $\hat{s}$ and $R_f$, the top $f\%$ of nodes ranked by the actual nodal influence $s$: 
  \begin{equation}
      Q_{r}{(\hat{s},s,f)} = \frac{\lvert R_f \cap V_f \rvert}{\lvert R_f \rvert}
  \end{equation}
  where $\lvert R_f \rvert$ = $f\%N$ is the number of nodes in $R_f$ and we take 20\% for $f\%$.

We evaluate the performance of each method via the average Kendall's correlation coefficient $\bar{Q}_k$ and the average Recognition rate $\bar{Q}_r$ over all possible starting times $t_0$ of the spreading process on each real-world network, as listed in \ref{Experimental settings}.

\section{Performance analysis}\label{Performance analysis}

We use the example of the HighSchool11 network to analyze the performance of the proposed influence prediction methods as similar key findings have been observed in the other real-world temporal networks. Results of the other networks are presented in the appendix. 

This Section is structured as follows: first, we systematically compare and explain the prediction quality of proposed centrality metrics. Next, we compare the prediction quality of centrality metrics that we proposed based on partial temporal network information with classical centrality metrics derived from full or partial (aggregated) network information. Finally, we evaluate the prediction quality of proposed methods in randomized real-world networks.

\subsection{Evaluating proposed metrics in real-world networks}\label{Prediction quality evaluation on real-world networks}

\begin{figure*}[h!]
  \centering
  \begin{minipage}[b]{1.0\textwidth}
    \subfigure[$\phi = 0.25$]{\includegraphics[width=\textwidth]{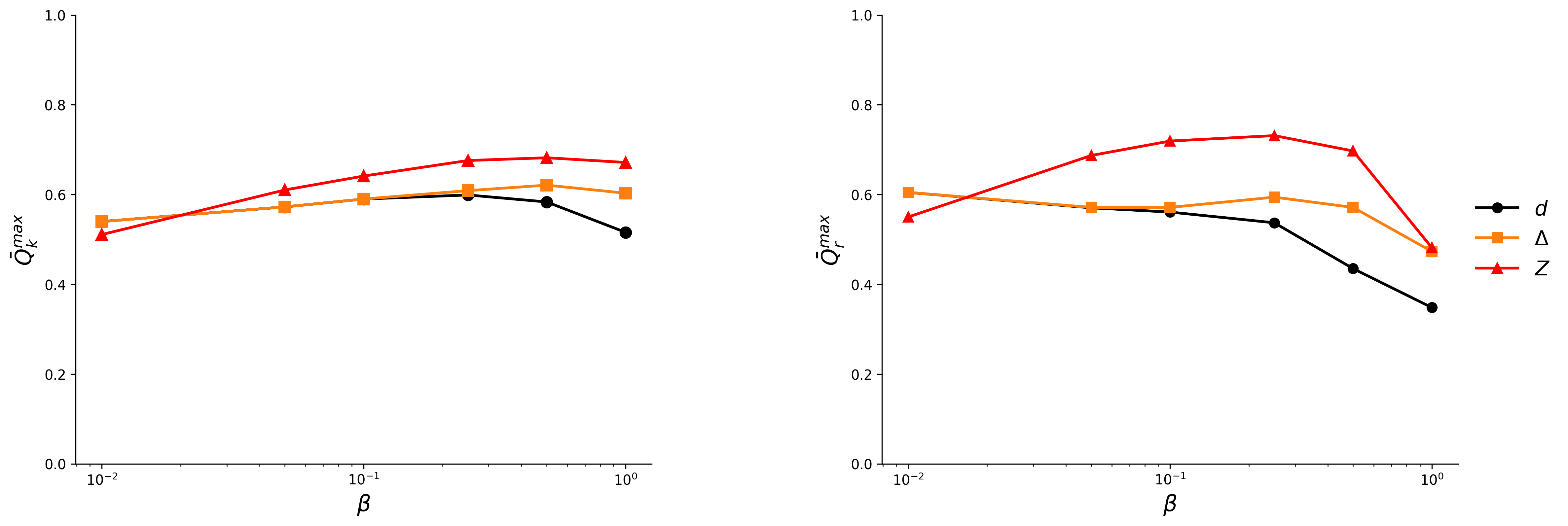}}
    \label{fig:subfig1}
  \end{minipage}
  \vspace{0.5cm}  % Optional: adds vertical space between the two images
  \begin{minipage}[b]{1.0\textwidth}
    \subfigure[$\phi = 0.5$]{\includegraphics[width=\textwidth]{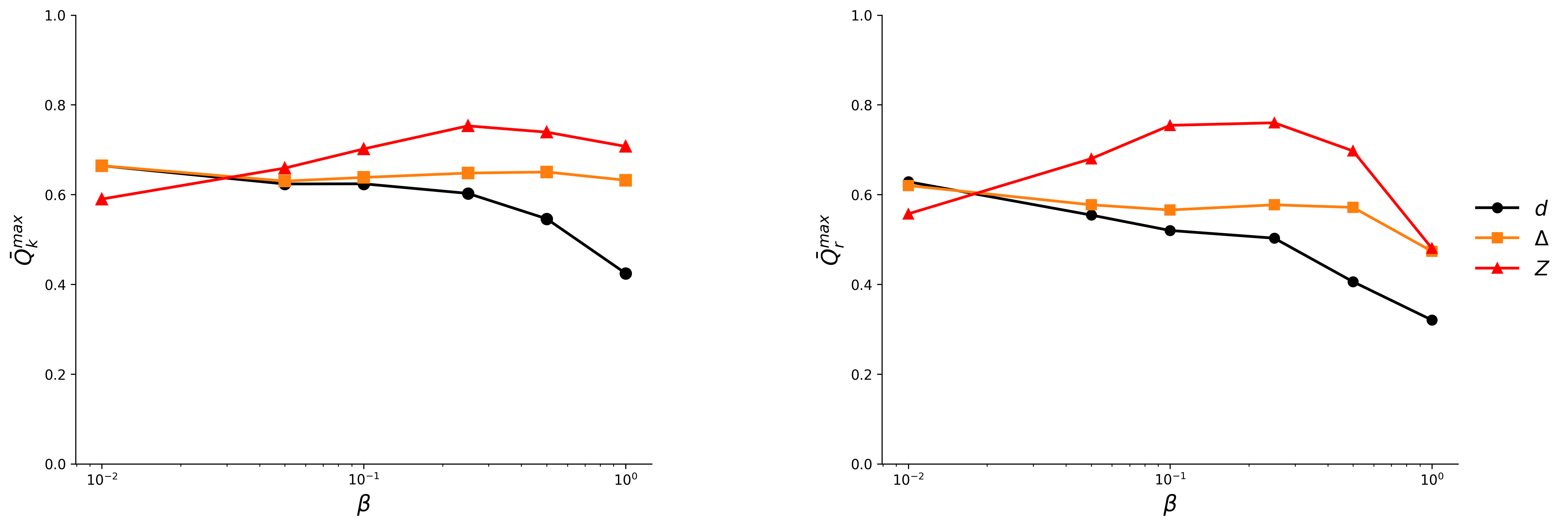}}
    \label{fig:subfig2}
  \end{minipage}
  \caption{The (best) prediction quality $\bar{Q}_k^{max}$ and $\bar{Q}_r^{max}$ of weighted degree mass ($d$), time-scaled temporal degree mass ($\Delta$), and time-scaled temporal reachability ($Z$), respectively, across various combinations of $\phi$ and $\beta$, in network HighScholl11.}
  \label{metrics performances}
\end{figure*}

First, we evaluate the prediction quality of the proposed centrality metrics, i.e., weighted degree mass ($d$), time-scaled temporal degree mass ($\Delta$), and time-scaled temporal reachability ($Z$) in estimating the ranking of nodes in influence. For each proposed centrality metric derived from partial network information and given a prediction quality evaluation measure Kendall’s correlation coefficient $\bar{Q}_k$ or Recognition rate $\bar{Q}_r$), the optimal parameter set $\{m, \alpha\}$ where $m \in [1,2,3]$ and $\alpha \in (0,1]$ that leads to the best quality will be considered and the corresponding best prediction quality is denoted as $\bar{Q}_k^{max}$ (or $\bar{Q}_r^{max}$). Figure \ref{metrics performances} shows how the (best) prediction quality of each proposed centrality metric varies with the relative duration $\phi$ of the partial temporal network and the infection probability $\beta$ of the diffusion process. When $\phi$ = 0.5, the proposed centrality metrics tend to achieve slightly better prediction quality compared with the case when $\phi = 0.25$, regardless of the value of $\beta$. Using temporal network information observed in a longer period, the proposed centrality metrics tend to estimate nodal influence better. 

We find that, in general, the time-scaled temporal reachability $Z$ performs the best whereas the weighted degree mass $d$ performs the worst when $\beta$ is relatively large. When $\beta$ is small, $Z$ performs the worst whereas the other two metrics performs equally well. These two observations can be explained as follows.

Considering the case when $\beta \rightarrow 0$. Each 1-hop neighbor of the seed node in the aggregated network $G^w$ could get infected with a probability $\beta$ times the number of contacts in between. The probability that a 2-hop neighbor gets infected is negligibly small, of order $\beta^2$. The influence of a node $i$ is proportional to the total number of contacts $d_i^{(1)}$ of the node in the temporal network over the complete period $[1, \tau]$, which is lower bounded by $d_i^{(\phi,1)}$, the weighted degree mass $d_i^{(\phi,1)}$ derived from the partial network information. Hence, weighted degree mass in the partial network is supposed to well predict nodal influence when $\beta \rightarrow 0$. When $\alpha=1$, $\Delta_i^{\phi,m=1}=\delta_i^{\phi,m=1}=d_i^{(\phi,1)}$. Hence, the time-scaled temporal degree mass $\Delta$ performs equally well as weighted degree mass $d$. 

\begin{figure}[htbp]
  \centering
  \includegraphics[width=0.5\textwidth]{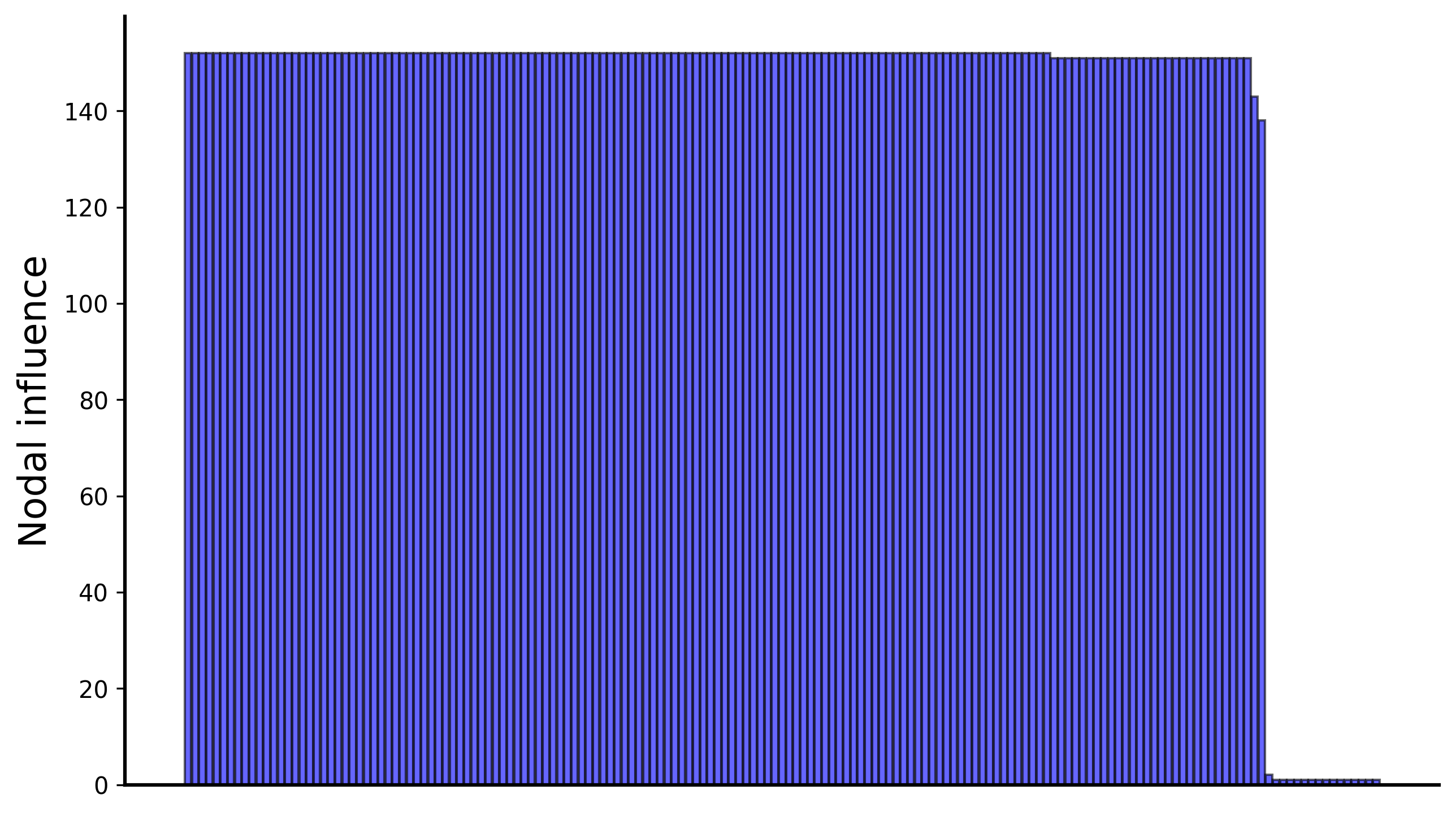} % 图片文件名为 example-image，可以替换为你的图片文件名
  \caption{The nodal influence of each node in a randomly selected observation period $[t_0+\tau]$ in the Manufacturing Emails network.}
  \label{The distribution of normalized nodal influence of $G$ during a randomly selected observation period in the Manufacturing Emails network.}
\end{figure}

When $\beta$ is large, a node $j$ further than 1-hop away from the seed node $i$ in the aggregated network $G^w$ could get infected. The corresponding spreading trajectory, i.e., the set of contacts via which $j$ gets infected, is a time-respecting path. Intuitively, the time-scaled temporal reachability $Z$ and time-scaled temporal degree mass $\Delta$, taking into account the time information of contacts, are supposed to perform better than the static metric $d$. When $\beta=1$, the actual influence of a node $i$ equals the number of distinct nodes that are reachable via time-respecting paths in full network $G$ starting from $i$ within $[1,\tau]$,  i.e., $Z_i^{(\phi=1,m \geq \rho)}$, where $\rho$ is the diameter of the aggregated network $G^w$, when $\alpha=1$. This supports why metric $Z$, i.e., $Z_i^{(\phi<1,m \leq 3)}$ which is smaller than $Z_i^{(\phi=1,m \geq \rho)}$, could perform the best when $\beta$ is large. An exception is observed in the Manufacturing Emails network, where $Z$ hardly outperforms other metrics when $\beta$ is large, as is shown in figure \ref{metrics performances of Manufacturing Emails}. The Manufacturing Emails network has the largest link density $p$ and the smallest modularity $\Gamma$ of the largest connected component of the weighted network $G^w$, as shown in table \ref{tab:real_world_networks}. As a consequence, the influence of most nodes approximately is equal to the size of the largest connected component in $G^w$ when $\beta=1$, as shown in figure \ref{The distribution of normalized nodal influence of $G$ during a randomly selected observation period in the Manufacturing Emails network.}. Hence, the influence of nodes are hardly distinguishable, such that the recognition rate of all proposed metrics is close to that of random guessing, i.e., 20\% when $\beta=1$. The similar nodal influence of most nodes supports why $Z$ hardly outperforms other metrics when $\beta$ is large.

\begin{figure}[htbp]
  \centering
  \includegraphics[width=0.5\textwidth]{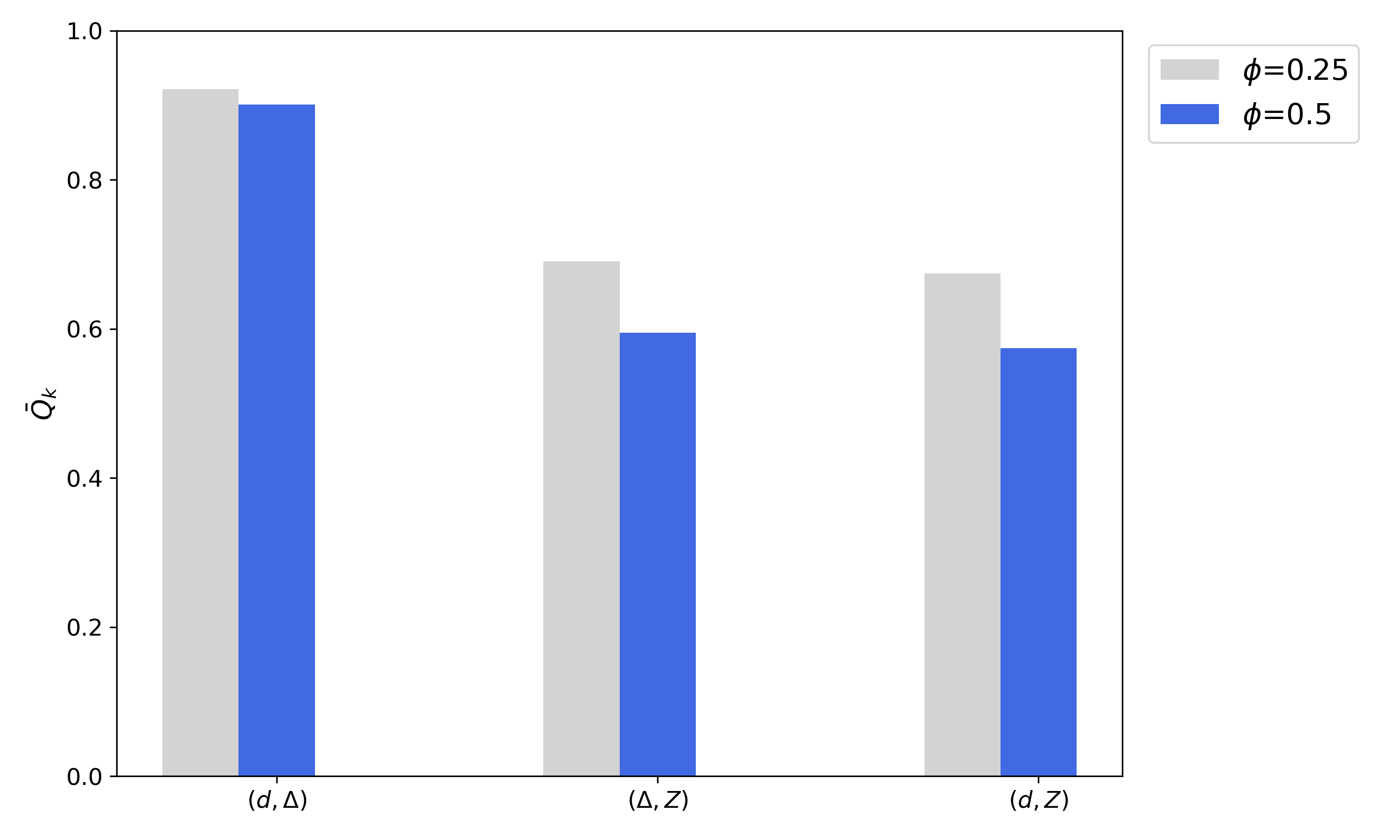} % 图片文件名为 example-image，可以替换为你的图片文件名
  \caption{Average Kendall correlation coefficient $\bar{Q}_k$ between every two proposed centrality metrics in Highschool11 with $m = 2$ and $\alpha = 1$ when $\phi$ = 0.25 and 0.5, respectively.}
  \label{kendall_hs11_metrics_correlation}
\end{figure}

\begin{figure}[htbp]
  \centering
  \includegraphics[width=0.5\textwidth]{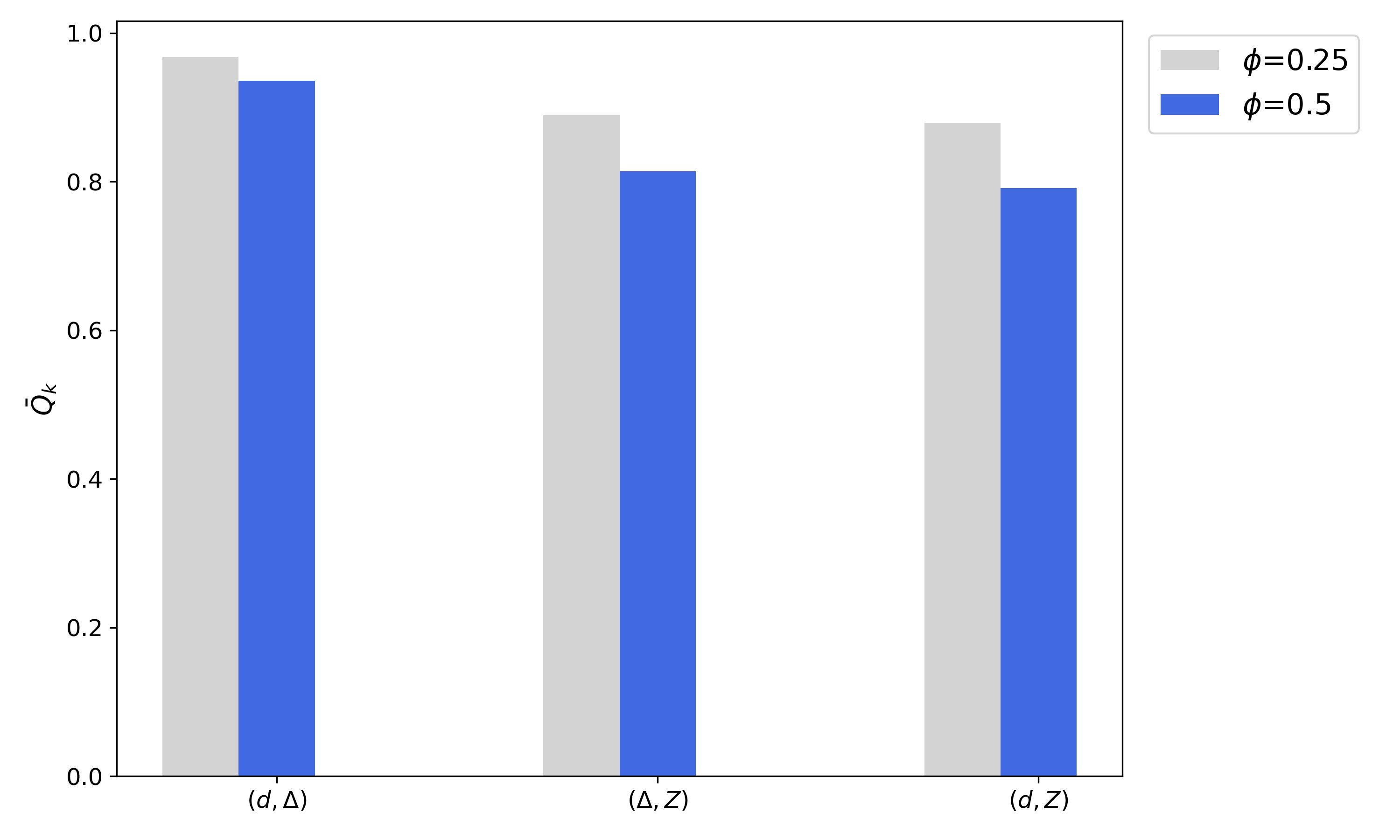} % 图片文件名为 example-image，可以替换为你的图片文件名
  \caption{Average Kendall correlation coefficient $\bar{Q}_k$ between every two proposed centrality metrics in Sms with $m = 2$ and $\alpha = 1$ when $\phi$ = 0.25 and 0.5, respectively.}
  \label{Average Kendall correlation coefficient between proposed centrality metrics in Sms}
\end{figure}

We explore further the similarity, i.e., Kendall's correlation between every pair of centrality metrics of a node derived from the same partial temporal network. This will help us understand whether these three proposed metrics capture similar properties of a node in the partially observed network and whether correlated metrics perform similarly in the predicting nodal influence or not. Consider $m = 2$ and $\alpha = 1$ as an example. Figure \ref{kendall_hs11_metrics_correlation} shows the average Kendall's correlation $\bar{Q}_k$ between every two metrics averaged over all possible starting times of the diffusion process, for $\phi = 0.25$ and $\phi = 0.5$, respectively.  It can be seen that all three centrality metrics are evidently and positively correlated. This is in line with their related definitions. This positive correlation $\bar{Q}_k<1$ also suggests that these metrics capture related but different properties of nodes. The weighted degree mass $d$ and time-scaled temporal degree mass $\Delta$ have the strongest correlation, which supports their similar prediction quality. Network Sms and Calls differ evidently from the other networks. In these two networks, the correlation between the weighted degree mass $d$ and time-scaled temporal degree mass $\Delta$ is significantly higher than that in other networks as exemplified in figure \ref{Average Kendall correlation coefficient between proposed centrality metrics in Sms}, and these two metrics perform almost the same in influence prediction (exemplified by \ref{metrics performances of Sms}). This high correlation between these two metrics in Sms and Calls can be explained as follows. Consider the case when $m=2$ as an example. The walks from a target node $i$ within $2$ hops on the aggregated partial networks include (1) 1-hop walks, (2) 2-hop walks that return to node $i$ and (3) 2-hop walks reaching finally other nodes than $i$. There are few type (3) walks, which is partially due to the low link density in the aggregated network of Sms and Calls (see Table \ref{tab:real_world_networks}) and supported by the low prevalence in these two networks (see Figure \ref{Average nodal influence in each real-world network}). When $\phi =1$, $d_i^{(\phi=1, m=2)}\approx\sum_{j}{[w(i,j)+w^2(i,j)]}$, where $w(i,j)$ is the total number of contacts between $i$ and $j$ in partial temporal $\mathcal{G}_{i}(\phi=1,m=2)$, or equivalently in full temporal network $G$. The time-respecting walks from $i$ within 2-hops on the partial temporal network include the same three types of walks. Hence, $\Delta_i^{(\phi=1, m=2)}\approx\sum_{j}{[w(i,j)+\frac{w(i,j)(w(i,j)-1)}{2}]}=\frac{1}{2}\sum_{j}{[w(i,j)+w^2(i,j)]}\approx \frac{1}{2}d_i^{(\phi=1, m=2)}$, when $\alpha=1$. Similarly, it holds that $\Delta_i^{(\phi, m=2)}\approx\frac{1}{2}d_i^{(\phi, m=2)}$ for any $\phi$ when $\alpha=1$.
%between a random node $A$ and other one node is the same as the number of 1-hop time-respecting walks between them, and the 2-hop walks/time-respecting walks from the node $A$ can be divided into two categories: (1) moving to node $A$'s neighbors and returning to node $A$ itself; (2) moving to other nodes except node $A$. It has been observed that, for both Sms and Calls, the sum of type (1) walks/time-respecting walks for a randomly selected node is generally greater than that of type (2) walks/time-respecting walks, which suggests the similarity between $d$ and $\Delta$ is dominated by the number of type (1) walks/time-respecting walks. Additionally, supposing node $A$ has $d$ neighbors and there are $n$ contacts between node $A$ and each of $d$ neighbors, then there will be totally $dn^2$ type (1) walks and $d\frac{n(n - 1)}{2} \approx d\frac{n^2}{2}$ type (1) time-respecting walks between node $A$ and other neighbors. The approximately linear mapping relation between $dn^2$ and $d\frac{n^2}{2}$ explains the reason why $d$ and $\Delta$ are strongly correlated.

% weighted degree mass ($d$), time-scaled temporal degree mass ($\Delta$), and time-scaled temporal reachability ($Z$)

\begin{figure*}[htbp]
  \centering
  \subfigure[$\bar{Q}_k$]{\includegraphics[width=0.48\textwidth]{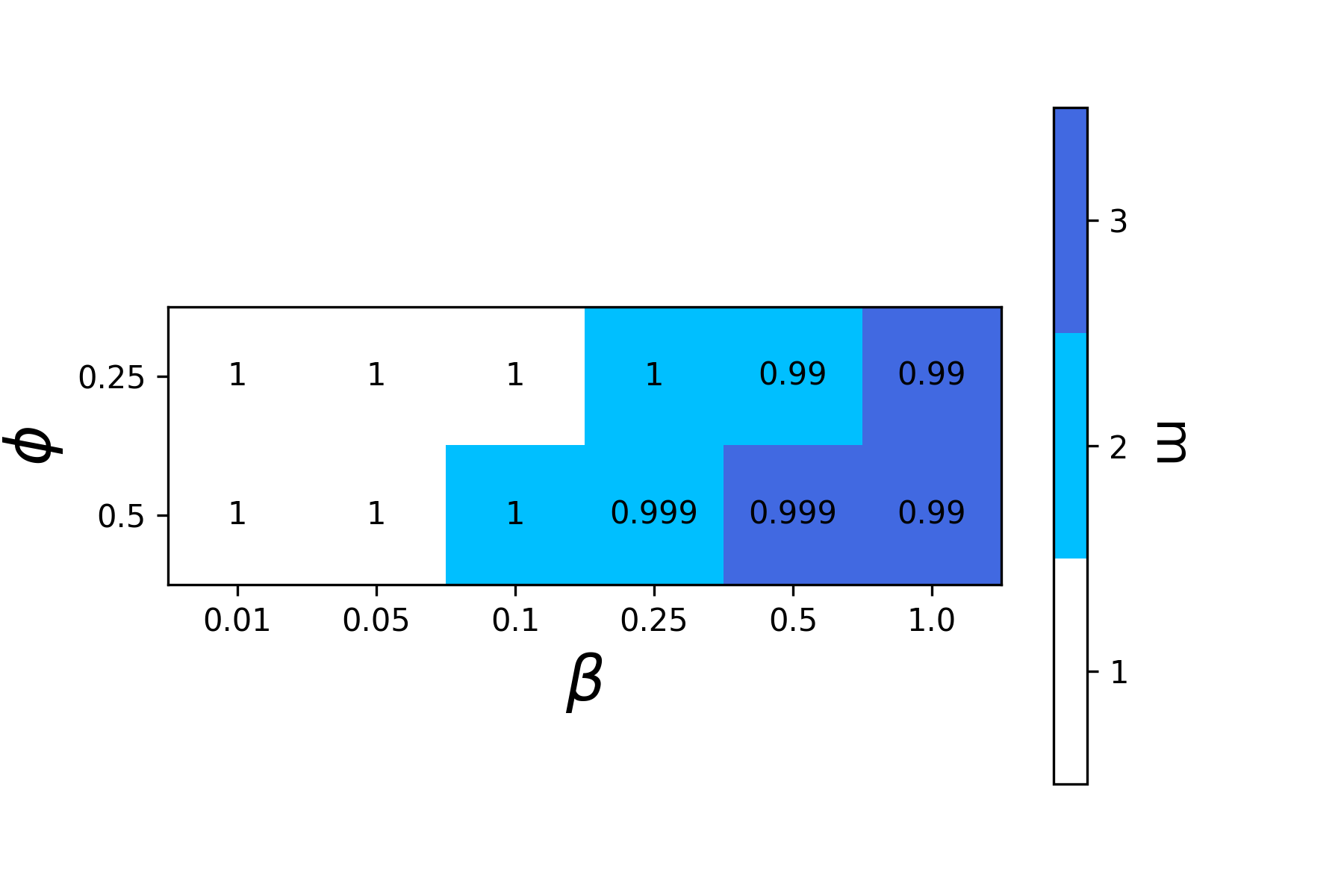}}
  \hfill
  \subfigure[$\bar{Q}_r$]{\includegraphics[width=0.48\textwidth]{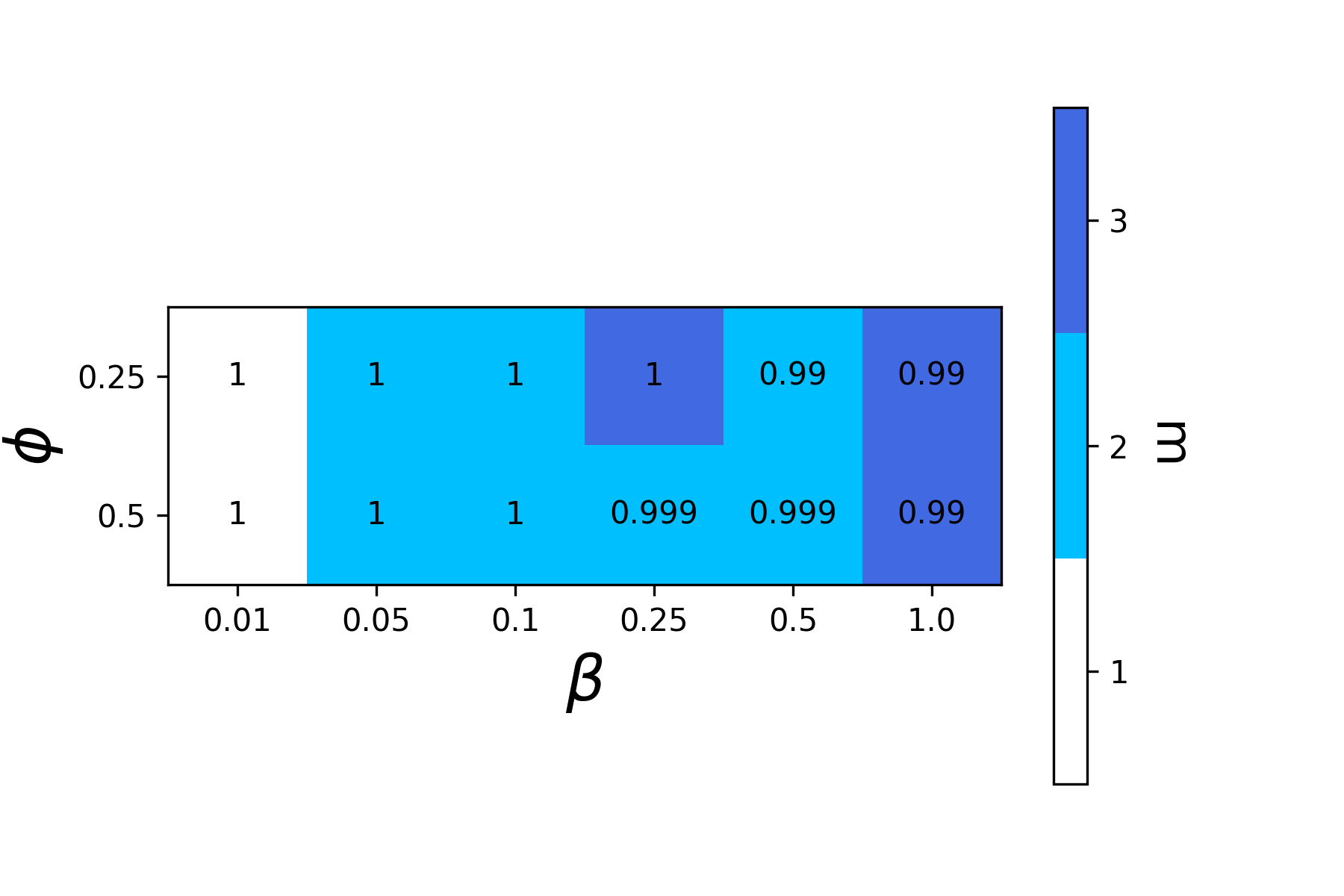}}
  \caption{The optimal $m$ and $\alpha$ that lead $Z$ to perform the best under different $\phi$ and $\beta$ combinations based on Highschool11 evaluated by $\bar{Q}_k$ and $\bar{Q}_r$, respectively. The numbers in the heapmap grid represent the optimal $\alpha$.}
  \label{A general observation of required hops of information}
\end{figure*}

Furthermore, we study the optimal parameter set $\{m, \alpha\}$ where $m \in [1,2,3]$ and $\alpha \in (0,1]$ for each centrality metric to achieve the best prediction quality $\bar{Q}_k^{max}$ and $\bar{Q}_r^{max}$, respectively. As time-scaled temporal reachability $Z$ performs the best except for the case when $\beta$ is small, we focus on metric $Z$. Figure \ref{A general observation of required hops of information} shows that when $\phi$ is fixed and as $\beta$ increases, the optimal $m$ increases. As discussed earlier, when $\beta \rightarrow 0$, the influence of a node is equal to the infection probability $\beta$ times the total number of contacts between this node and its 1-hop neighbors over $[t_0,t_0+\tau]$. When $\beta \rightarrow 0$, the information can hardly diffuse to a node that is further than 1-hop away from the seed node. Hence, only the partial network within 1-hop is relevant for influence prediction. As $\beta$ increases, the partial network connecting nodes further than 1-hop from the seed could possibly spread the information, thus relevant for prediction.

%o optimal prediction quality of $Z$. The reason could be that 3-hop neighbors of the seed node cannot be infected due to the low infection probability. But as $\beta$ becomes large, using 3-hop topology information can better predict how many nodes are infected. This can attributed to when the infection probability is large, 3-hop neighbors of the seed node can be very likely to be infected. In this case, 3-hop topology information is more relevant for predicting infected node numbers. 
Another finding is that when $\phi$ is fixed and as $\beta$ increases, the optimal\footnote{The optimal $\alpha$ is found via searching within \{0.2, 0.4, 0.6, 0.8, 0.85, 0.9, 0.95, 0.99, 0.999, 0.9999, 1\} decreases. Since it may take a long time to reach a node via the fastest time-respecting path, the metric $Z$ with $\alpha=0.9999$ and $\alpha=1$ respectively could differ evidently in prediction quality.} $\alpha$ decreases from $\alpha=1$. When $\beta$ is large, the time-scaled temporal reachability $Z$ is supposed to perform better than the unscaled case, i.e., $\alpha=1$ because a node tends to be influential if it can reach many nodes and each reachable node is reached early enough for it to potentially infect other nodes further. However, this is not the case when $\beta$ is small, because nodes further than 1-hop away from the seed can hardly get infected. In networks Sms and Calls which has a low link density in the aggregated network, nodes further than 1-hop away from the seed can hardly get infected either, even when $\beta$ is large. Hence, $\alpha=1$ leads to the best performance of $Z$ in influence estimation, independent of $\beta$, in these two networks. 

%As for $\alpha$, we have observed that the scaled $Z$ typically outperforms the unscaled $Z$ when $\beta$ is large, whereas the unscaled $Z$ generally exhibits superior prediction quality compared to the scaled $Z$ when $\beta$ is small. The reason for this is the seed node will possibly infect other nodes through faster reachable nodes when the infection probability is large. In this scenario, the scaled $Z$, which picks out faster reachable nodes through assigning more weight, is more concerned with the number of infected nodes. On the contrary, when the infection probability is small, the number of nodes that a seed node can infect primarily depends on the number of time-respecting walks between it and other nodes, while the time it takes to reach these nodes is no longer significant. Therefore, unscaled $Z$, which counts the number of reachable nodes (positively correlated with the number of time-respecting walks), demonstrates superior prediction quality. A similar tendency is also observed for $d$ and $\Delta$ (for $d$, scaling is not involved, the varying trend of $m$ with $\beta$, independent of $\phi$, exhibits a similar pattern). Note that the duration of the time-respecting walk and the reached time of the reachable node reflect the potential of the seed node to spread epidemics to further neighbors in the context of only partial temporal network information is available. If complete temporal network information is provided, it is meaningless to scale centrality metrics. 

\subsection{Comparison with classic centrality metrics in real-world networks} \label{Comparison with classic centrality metrics in prediction quality}

\begin{figure*}[htbp]
  \centering
  \includegraphics[width=\textwidth]{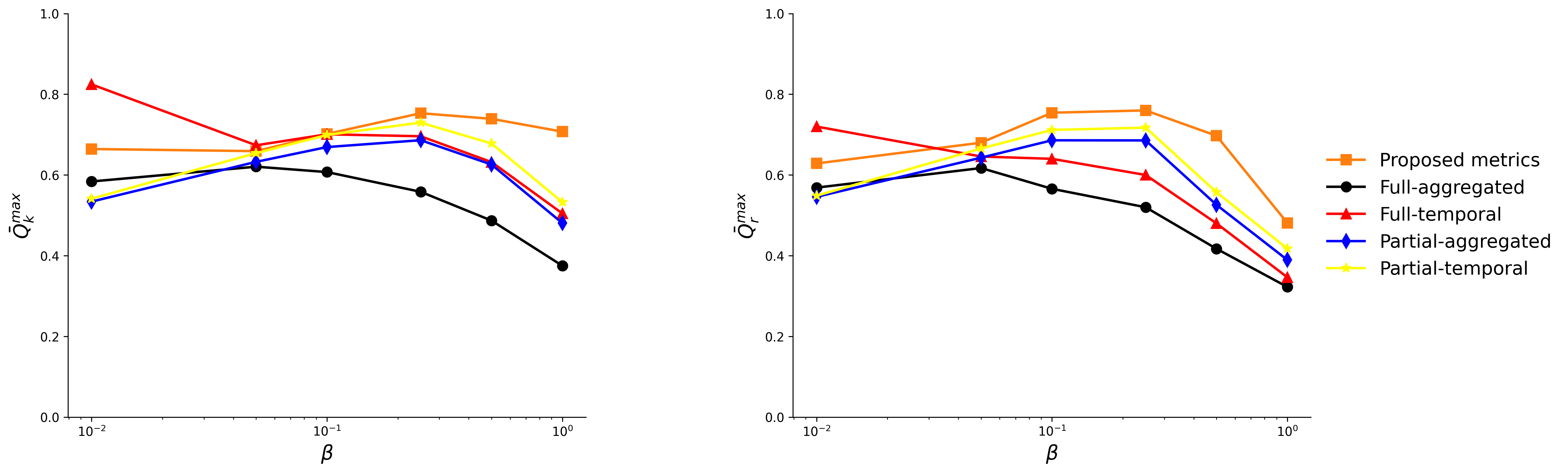}
  \vspace{2mm} % 增加间距，避免文字紧贴图像
  \text{(a) $\phi = 0.5$} % 单独添加 $\phi$ 参数下标
  \caption{
  The best prediction quality $\bar{Q}_k^{max}$ and $\bar{Q}_r^{max}$, respectively achieved by proposed centrality metrics derived from the partial temporal network $\mathcal{G}_{i}(\phi,m)$ (denoted by orange squares); full-aggregated: each static centrality metric derived from the unweighted aggregated network of the full temporal network $G$ (denoted by black dots); full-temporal: the average of each static centrality metric derived from all snapshots of $G$ or the temporal closeness centrality derived from $G$ (denoted by red triangles); partial-aggregated: each static centrality metric derived from the unweighted aggregated network of the partial temporal network $\mathcal{G}_{i}(\phi,m)$ (denoted by blue diamond); partial-temporal: the average of each static centrality metric derived from all snapshots of the partial temporal network $\mathcal{G}_{i}(\phi,m)$ or temporal closeness centrality derived from $\mathcal{G}_{i}(\phi,m)$ (denoted by yellow stars) when $\phi=0.5$ and $\beta$ varies, based on Highschool11.
  }
  \label{Prediction performance comparison between proposed local centrality metrics and classic global centrality metrics}
\end{figure*}

We are interested in how these proposed metrics derived from partial temporal network information perform, in comparison with classic centrality metrics, defined in \ref{Centrality metrics based on full temporal network}, utilizing full or partial, temporal or aggregated network information. Classic centrality metrics (4 static metrics and temporal closeness) can be derived for influence prediction via 4 possible ways: (1) Full-aggregated: each static centrality metric derived from the unweighted aggregated network of the full temporal network $G$ (2) Full-temporal: the average of each static centrality metric derived from all snapshots of $G$ or the temporal closeness centrality derived from $G$ (3) Partial-aggregated: each static centrality metric derived from the unweighted aggregated network of the partial temporal network $\mathcal{G}_{i}(\phi,m)$ or (4) Partial-temporal: the average of each static centrality metric derived from all snapshots of the partial temporal network $\mathcal{G}_{i}(\phi,m)$ or temporal closeness centrality derived from $\mathcal{G}_{i}(\phi,m)$. Given the three parameters coming from the context of the prediction problem: $\beta$, $\phi$, and the measure to evaluate the prediction quality, we compare the best performance achieved by each of the four classes of methods using classic metrics and the best performance achieved by our proposed centrality metrics in figure \ref{Prediction performance comparison between proposed local centrality metrics and classic global centrality metrics}.

\begin{figure}[htbp]
  \centering
  \includegraphics[width=0.5\textwidth]{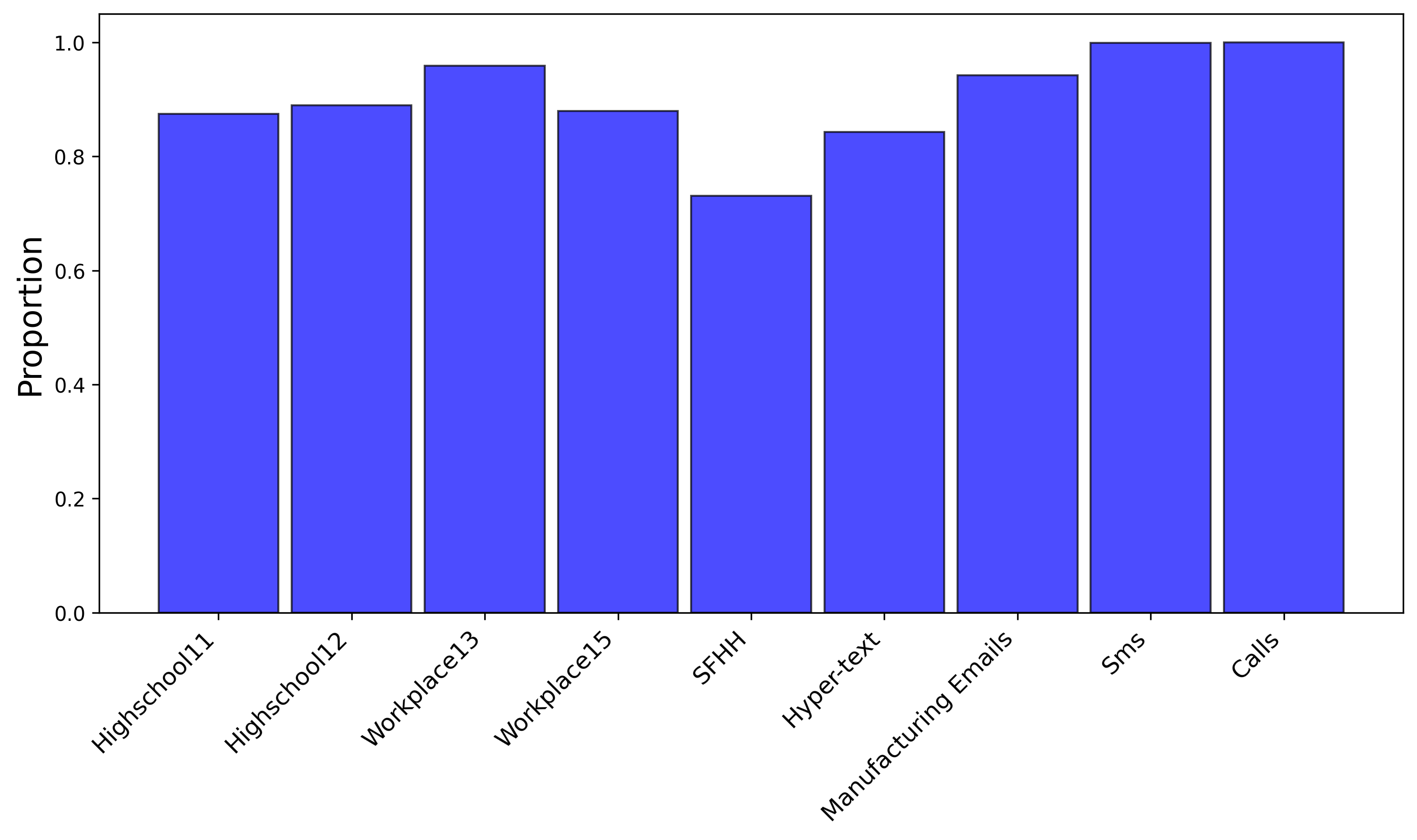} 
  \caption{Proportion of nodes that have degree 1 among nodes that have any contact in a snapshot, averaged over all snapshots of each real-world network.}
  \label{The proportion of nodes who have degree 1 over all snapshots of 9 considered real-world networks.}
\end{figure}

As shown in figure \ref{Prediction performance comparison between proposed local centrality metrics and classic global centrality metrics}, either our proposed metrics or classic metrics derived from full temporal networks perform the best. This indicates that introducing temporal information in the design of centrality metrics could be beneficial. The centrality metrics we have proposed perform mostly the best. This suggests that those walk-related properties of a node in a partial network well indicate the influence of the node in the global network in the long term. It is found that when $\beta \rightarrow 0$, the average closeness over all snapshots of a temporal network $G$ generally performs the best. This can be explained as follows: it is observed that in each snapshot of each temporal network, most contacts are disjoint; This is supported by high proportion of nodes that have degree 1 among nodes that have any contact in each snapshot, as illustrated in figure \ref{The proportion of nodes who have degree 1 over all snapshots of 9 considered real-world networks.}; Hence, the average closeness of a node over all snapshots of $G$ approximates the average degree of this node over all snapshots, proportional to the total number of contacts of the node over all snapshots; As explained in Section \ref{Prediction quality evaluation on real-world networks}, the influence of a node when $\beta \rightarrow 0$ is equal to the total number of contacts of the node in $G$; Hence, when $\beta \rightarrow 0$, the average closeness over all snapshots tends to provide the best estimate of nodal influence.
%[read carefully to understand why the text below can not be used]Hence, the average closeness of a node over all snapshots of $G$, which approximates the average degree of this node over all snapshots, is proportional to the number of infected nodes from this node as the seed node when $\beta \rightarrow 0$ on $G$ as the infected nodes are mostly 1-hop neighbors of the seed node in the aggregated network of $G$ (the reason why infected nodes are mostly 1-hop neighbors of the seed node when $\beta \rightarrow 0$ was discussed in \ref{Prediction quality evaluation on real-world networks}). 
%把为什么Manufacturing Emails network特殊的解释先去掉，后面再说。
%However, in the Manufacturing Emails network, star networks\footnote{A star network refers to a type of network architecture where all peripheral nodes are connected to a central node and no direct links exist between peripheral nodes.} are more frequently observed in each snapshot of the temporal network. These star networks lead to the fact that for a peripheral node in the star network, the average closeness of the node over all snapshots of $G$ does not approximate the average degree of this node over all snapshots, as 2-hop paths are considered. This, in turn, causes the average closeness over all snapshots of the node to predict worse than $\Delta$ and $d$ when $\beta$ is small (the reason why $\Delta$ and $d$ could perform equally well when $\beta$ is small was discussed in \ref{Prediction quality evaluation on real-world networks}). 

\begin{figure}[htbp]
  \centering
  \includegraphics[width=0.5\textwidth]{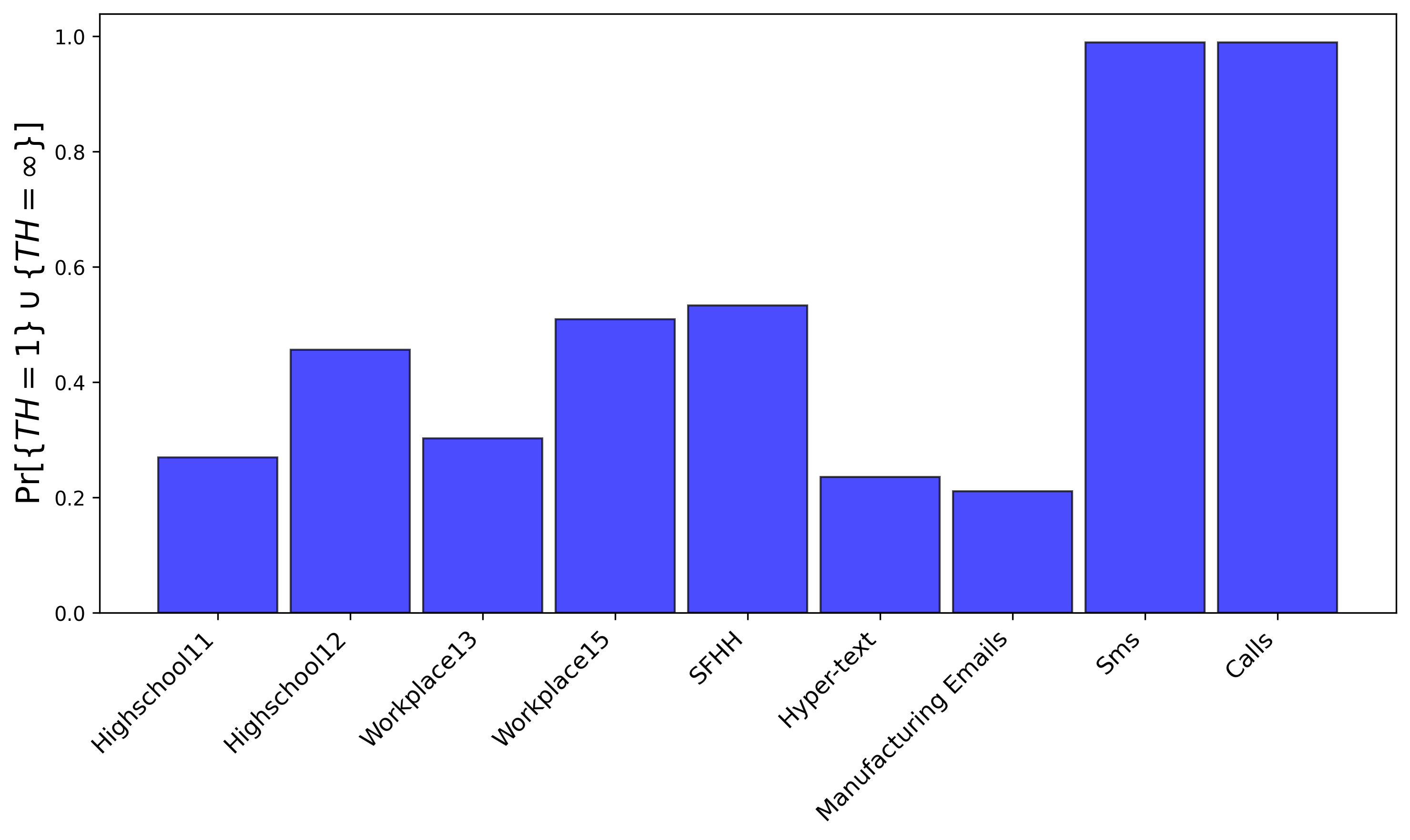} 
  \caption{Probability $Pr[\{TH=1\}\cup\{TH=\infty\}]$ that the hopcount of the shortest time-respecting path from a random node to another random node is either 1 or infinity, averaged over all observation periods $t_0+\tau$ of each real-world network.}
  \label{The proportion of hopcounts equal to 1 or infinity}
\end{figure}

When $\beta$ is large, the class of metrics we have proposed, actually time-scaled temporal reachability $Z$ derived from the partial temporal network generally performs the best among all metrics. However, in the network Sms and Calls, temporal closeness centrality derived from $G$ performs the best, when $\beta$ is large. This can be attributed to the following reasons. Consider the case when $\beta=1$. The influence of a target node in a temporal network is equal to the number of distinct nodes that are reachable by the target node via the shortest time-respecting paths. If the hop-count of the shortest time-respecting path from the target node to any node is either 1 or infinity (meaning a time-respecting path from the target to this node does not exist), the temporal closeness of the target node is equal to its influence. Figure \ref{The proportion of hopcounts equal to 1 or infinity} shows that the probability $Pr[\{TH=1\}\cup\{TH=\infty\}]$ that the hopcount of the shortest time-respecting path from a random node to another random node is either 1 or infinity is the largest, close to 1, in Sms and Calls, explaining why temporal closeness perform the best in these two networks, when $\beta$ is large. This probability $Pr[\{TH=1\}\cup\{TH=\infty\}]$ is large in Sms and Calls because these two networks have a low link density $p$ and a large average modularity $\Gamma$ of the largest connected component of the weighted aggregated network $G^w$, as reflected in table \ref{tab:real_world_networks}. Nodes mainly interact within small communities, such that a node pair is likely either not connected via a time-respecting path when they belong to different communities or connected via a time-respecting path with a small hopcount when they belong to the same community.

%in these two networks, due to the low link density $p$ as reflected in table \ref{tab:real_world_networks} and the large average of modularity (Sms: 0.928; Calls: 0.935) of the largest connected component of the weighted network $G^w$ across all considered observation periods, the influence of most nodes is relatively small and varies significantly when $\beta$ is large. Additionally, for the same two factors mentioned above — the sparse distribution of links between nodes and the segregated community structure — the hop-count values of the shortest time-respecting paths between nodes are mostly equal to 1 or infinity, as shown in the figure \ref{The distribution of hopcounts equal to 1 or infinity}. When the hop-count values of the shortest time-respecting paths between nodes are either 1 or infinite, the rank correlation between temporal closeness and nodal influence will be 1, provided that nodal influence varies significantly. This explains why temporal closeness performs noticeably better than other metrics when $\beta$ is not small, as the rank of temporal closeness approximates the rank of nodal influence.}

% \begin{figure*}[htbp]
%   \centering
%   % First row: 5 figures
%   \subfigure[Workplace13]{\includegraphics[width=0.19\textwidth]{figures/Hopcountwork1_plot.png}}
%   \hfill
%   \subfigure[Workplace15]{\includegraphics[width=0.19\textwidth]{figures/Hopcountwork2_plot.png}}
%   \hfill
%   \subfigure[Highschool11]{\includegraphics[width=0.19\textwidth]{figures/Hopcounths11_plot.png}}
%   \hfill
%   \subfigure[Highschool12]{\includegraphics[width=0.19\textwidth]{figures/Hopcounths12_plot.png}}
%   \hfill
%   \subfigure[Hyper-text]{\includegraphics[width=0.19\textwidth]{figures/Hopcounthyper_text_plot.png}}
  
%   \vspace{0.5cm} % Adjust space between rows
  
%   % Second row: 4 figures aligned with the first 4 figures above
%   \subfigure[SFHH]{\includegraphics[width=0.19\textwidth]{figures/Hopcountscientific_conference_plot.png}}
%   \hfill
%   \subfigure[Manufacturing Emails]{\includegraphics[width=0.19\textwidth]{figures/Hopcountmanufacturing_emails_plot.png}}
%   \hfill
%   \subfigure[Sms]{\includegraphics[width=0.19\textwidth]{figures/Hopcountfiltered_sms_plot.png}}
%   \hfill
%   \subfigure[Calls]{\includegraphics[width=0.19\textwidth]{figures/Hopcountfiltered_calls_plot.png}}
%   \hspace*{\fill} % Add extra horizontal space to align properly
  
%   \caption{The distribution of hopcounts for the shortest time-respecting paths between nodes across all observation periods of 9 considered real-world networks.}
%   \label{fig:hopcount_distribution}
% \end{figure*}

In addition to evaluating the performance of the proposed centrality metrics in terms of predication quality, we briefly address their computational complexity. Let $\mathcal{E}$ represent the total number of contacts occurring during the partial observation period [1,$\phi\tau$], and suppose the maximum number of contacts for any given node during this period is $\mathcal{L}$. The computational complexity of weighted degree mass $d$, time-scaled temporal degree mass $\Delta$, and time-scaled temporal reachability $Z$ for all nodes derived from the partial temporal network $\mathcal{G}_{i}(\phi,m)$ is shown in table \ref{computational complexity}. The polynomial complexity of two metrics poses challenges in computing them in full temporal networks.

\begin{table}[h]
\centering
\caption{Complexity of computing each proposed centrality metric for all nodes based on the partial temporal network $\mathcal{G}_{i}(\phi,m)$.}
\label{computational complexity}
\resizebox{\textwidth}{!}{% 调整表格大小为文本宽度
\begin{tabular}{ccc}
\toprule
 $d$ & $\Delta$ & $Z$  \\
\midrule
$\mathcal{O}(\mathcal{E}m)$ & $\mathcal{O}(N\mathcal{L}^m)$ & $\mathcal{O}(N\mathcal{L}^m)$ \\
\bottomrule
\end{tabular}
}
\end{table}

\subsection{Evaluating proposed metrics in randomized real-world networks} \label{Prediction quality evaluation on randomized networks}

The performance of an influence prediction method may depend on the properties of the underlying temporal network. The contact networks we considered manifest correlations between contacts. For example, contacts that are close in topology tend to be close in time. Hence, we explore further whether our findings in the performance of proposed metrics and of the classic metrics are still valid when these real-world networks are randomized, i.e., the correlation between topology and time are removed. 
%Section \ref{Prediction quality evaluation on real-world networks} explores how the prediction quality of proposed centrality metrics is influenced by the hops or time duration of walks/time-respecting walks in the partially observed temporal network, essentially the topological and temporal information of contacts within the network. Additionally, we're interested in whether the interrelation between topological and temporal information of contacts could affect the prediction quality of our metrics. To explore this, the temporal network randomization technique introduced by \citep{ceria2022topological} is employed. 
Consider the set of contacts ${l(i, j,t)}$ in a temporal network $G$, where each contact is described by its topological location, i.e., between pair of nodes $(i, j)$ and its time stamp t. A randomized network of $G$ is obtained by reshuffling the time stamps of contacts in the network, without changing the topological locations of contacts \citep{ceria2022topological}. This randomization does not change the number of contacts between each node pair. Only the time stamps of contacts are randomly switched. 

\begin{figure*}[h!]
  \centering
  \begin{minipage}[b]{1.0\textwidth}
    \subfigure[$\phi = 0.25$]{\includegraphics[width=\textwidth]{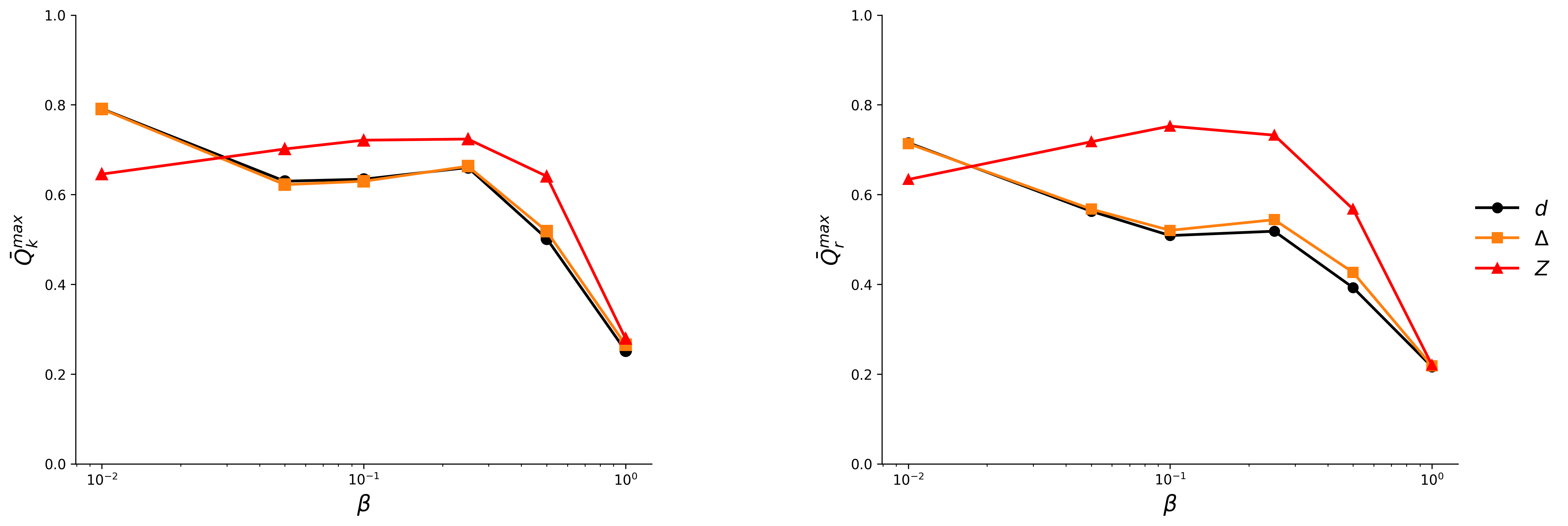}}
    \label{fig:subfig1}
  \end{minipage}
  \vspace{0.5cm}  % Optional: adds vertical space between the two images
  \begin{minipage}[b]{1.0\textwidth}
    \subfigure[$\phi = 0.5$]{\includegraphics[width=\textwidth]{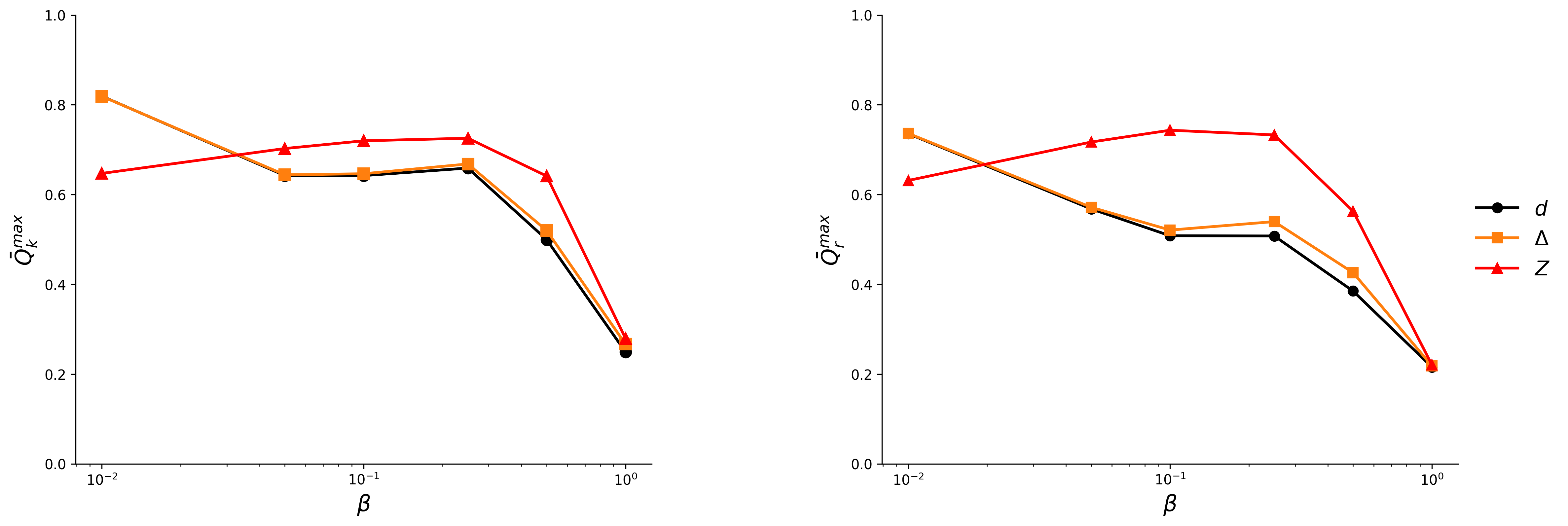}}
    \label{fig:subfig2}
  \end{minipage}
  \caption{The optimal prediction quality $\bar{Q}_k^{max}$ and $\bar{Q}_r^{max}$ of weighted degree mass ($d$), time-scaled temporal degree mass ($\Delta$), and time-scaled temporal reachability ($Z$) across various $\phi$ and $\beta$, averaged on 10 randomized networks of Highschool11.}
  \label{metrics on randomized networks}
\end{figure*}

As shown in figure \ref{metrics on randomized networks}, we observe the same result in real-world networks and in their randomized networks that the time-scaled reachability performs the best for a broad range of $\beta$ that is not small and the other two proposed metrics perform the best when $\beta$ is small. A different observation in randomized networks is that when $\beta=1$, all proposed metrics perform similarly and badly. This is because the spreading influence of all nodes in the randomized network tends to be higher than that in the original network and tends to be identical when $\beta=1$. For example, any seed node may lead to the infection of all the other nodes within the same connected component within $[1,\tau]$. The influence of nodes are hardly distinguishable in this case, similar to the case when $\beta$ is large in the Manufacturing Emails network. Meanwhile, it is observed that the Kendall's correlation between weighted degree mass $d$ and time-scaled temporal degree mass $\Delta$ gets stronger after network randomization when $\alpha=1$ and $m \in \{2, 3\}$. Intuitively, a node that has a large weighted degree mass $d$, i.e., many walks within m hops in the aggregated network of the partial network tends to have a large temporal degree mass, i.e., many time-respecting walks within m hops, because the time stamps are assigned randomly.
%the probability of any two time stamps on adjacent contacts being in either increasing or decreasing order is \( \frac{1}{2} \). Suppose in a walk $(n_0,n_1),(n_1,n_2),..(n_k-1,n_k)$, the time stamp increasing condition for any two adjacent contacts is an independent event. Then, the probability that the entire walk satisfies the time stamps increasing condition (i.e., being a time-respecting walk) is $\left( \frac{1}{2} \right)^{k-1}$, where \( k \) is the number of contacts in the walk. The overlapping probability \(P(\text{k-hop time-respecting walks}  \cap  \text{k-hop walks})\) between node $n_0$ and $n_k$ can be expressed as the ratio of the number of k-hop time-respecting walks to the total number of k-hop walks and the expression is 

\begin{figure*}[htbp]
  \centering
  \includegraphics[width=\textwidth]{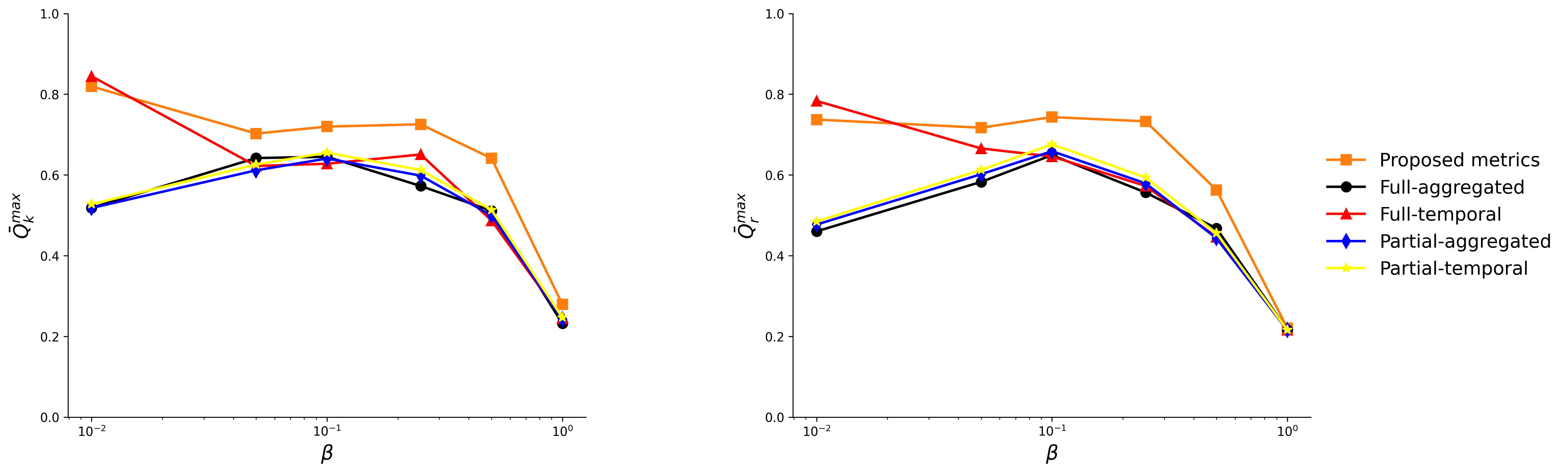}
  \vspace{2mm} % 增加间距，避免文字紧贴图像
  \text{(a) $\phi = 0.5$} % 单独添加 $\phi$ 参数下标
  \caption{The best prediction quality $\bar{Q}_k^{max}$ and $\bar{Q}_r^{max}$, respectively achieved by proposed centrality metrics derived from the partial temporal network $\mathcal{G}_{i}(\phi,m)$ (denoted by orange squares); full-aggregated: each static centrality metric derived from the unweighted aggregated network of the full temporal network $G$ (denoted by black dots); full-temporal: the average of each static centrality metric derived from all snapshots of $G$ or the temporal closeness centrality derived from $G$ (denoted by red triangles); partial-aggregated: each static centrality metric derived from the unweighted aggregated network of the partial temporal network $\mathcal{G}_{i}(\phi,m)$ (denoted by blue diamond); partial-temporal: the average of each static centrality metric derived from all snapshots of the partial temporal network $\mathcal{G}_{i}(\phi,m)$ or temporal closeness centrality derived from $\mathcal{G}_{i}(\phi,m)$ (denoted by yellow stars) when $\phi=0.5$ and $\beta$ varies, averaged on 10 randomized networks of Highschool11.}
  \label{Prediction performance comparison between proposed local centrality metrics and classic global centrality metrics on randomized networks}
\end{figure*}

Furthermore, we compare the prediction quality of proposed metrics and classic metrics in the randomized real-world networks, as exemplified in figure \ref{Prediction performance comparison between proposed local centrality metrics and classic global centrality metrics on randomized networks}. The same result has been found that our proposed centrality metric set performs mostly better than classic centrality metrics derived from both the full and partial temporal network, as well as their corresponding unweighted aggregated networks. The performance of our metrics tends to be robust to variations in the correlation between temporal and topological information of contacts.

\section{Conclusions and future work}\label{Conclusions and future work}

In this work, we address the problem of using partial temporal network information, i.e., local network observed over a short period, to predict nodal spreading influence on the full temporal network over a long period. This study also aims to identify which network properties of a node in the partial network determine this node's influence. 
The spreading influence of a node depends on how well the node is connected to other nodes via possible spreading trajectories. The spreading trajectory from any seed node to any other node in a temporal network is not necessarily the shortest time-respecting path, but can be any time-respecting path. This motivates us to design centrality metrics that systematically capture how well a node is connected in the partial network via (time-respecting) walks. These metrics contrast with class metrics describing the connection of a node to other nodes via the shortest (time-respecting) paths. The quality of these metrics in estimating the ranking of nodes in influence is evaluated and compared against classical centrality metrics in real-world contact networks and their randomized networks. We find and explain that the proposed metrics using the partial network mostly outperform classic centrality metrics derived from the full temporal network, across a broad range of the infection probability. A node tends to be influential if it can reach many distinct nodes via time-respecting walks and if these nodes can be reached early in time. 

This study has several limitations that call for further exploration. Firstly, this work focuses on estimating the influence of nodes in the SI process on temporal networks. The proposed methods can be applied and extended for other spreading processes. Additionally, epidemics and information may spread via higher-order or group interactions \citep{wang2024epidemic,battiston2020networks,aktas2021identifying}. It is interesting to explore the feasibility of predicting nodal influence in higher-order temporal networks using partial network information. Secondly, how the parameters $m$ and $\alpha$ of the proposed centrality metrics affect the prediction quality has been analyzed and explained. It is intriguing to investigate whether certain universal ranges of these parameters tend to lead to near-optimal estimation quality for certain types of networks. Finally, the polynomial computational complexity of the proposed metrics limits their scalability in large-scale temporal networks. Given their promising prediction quality, designing efficient approximation algorithms is worthwhile.
%are empirically selected under different experimental settings, as well as the parameters for classic centrality metrics. These selected parameter values might be suboptimal, and therefore identifying the optimal parameters for practical applications remains an open question that requires further exploration. 

\section{Data availability}
The data sets used are publicly available. More information can be found in the corresponding references.

\section{Acknowledgements}
We thank for the support of Netherlands Organisation for Scientific Research NWO (project FORT-PORT no. KICH1.VE03.21.008) and the China Scholarship Council (CSC). We greatly appreciate Dr. Márton Karsai for his valuable comments and suggestions.

\newpage
\appendix
\section{Prediction quality of proposed metrics in other real-world networks} 
\fontsize{11pt}{12pt}\selectfont

\begin{figure*}[h!]
  \centering
  \begin{minipage}[b]{1.0\textwidth}
    \subfigure[$\phi = 0.25$]{\includegraphics[width=\textwidth]{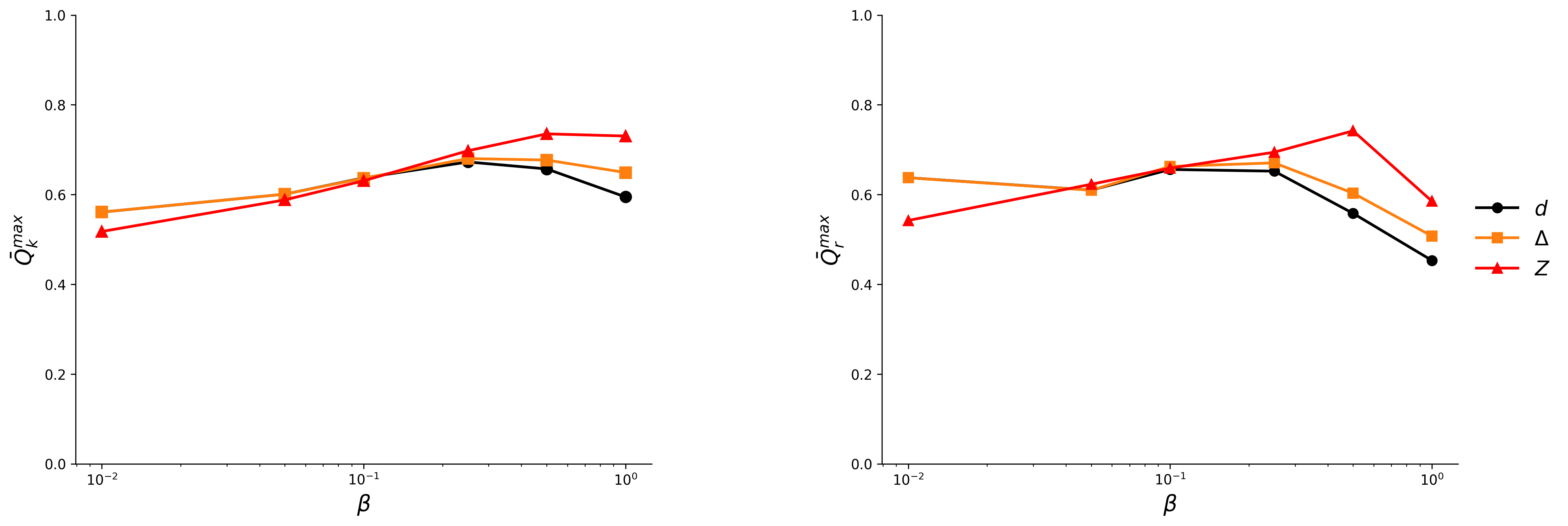}}
    \label{fig:subfig1}
  \end{minipage}
  \vspace{0.5cm}  % Optional: adds vertical space between the two images
  \begin{minipage}[b]{1.0\textwidth}
    \subfigure[$\phi = 0.5$]{\includegraphics[width=\textwidth]{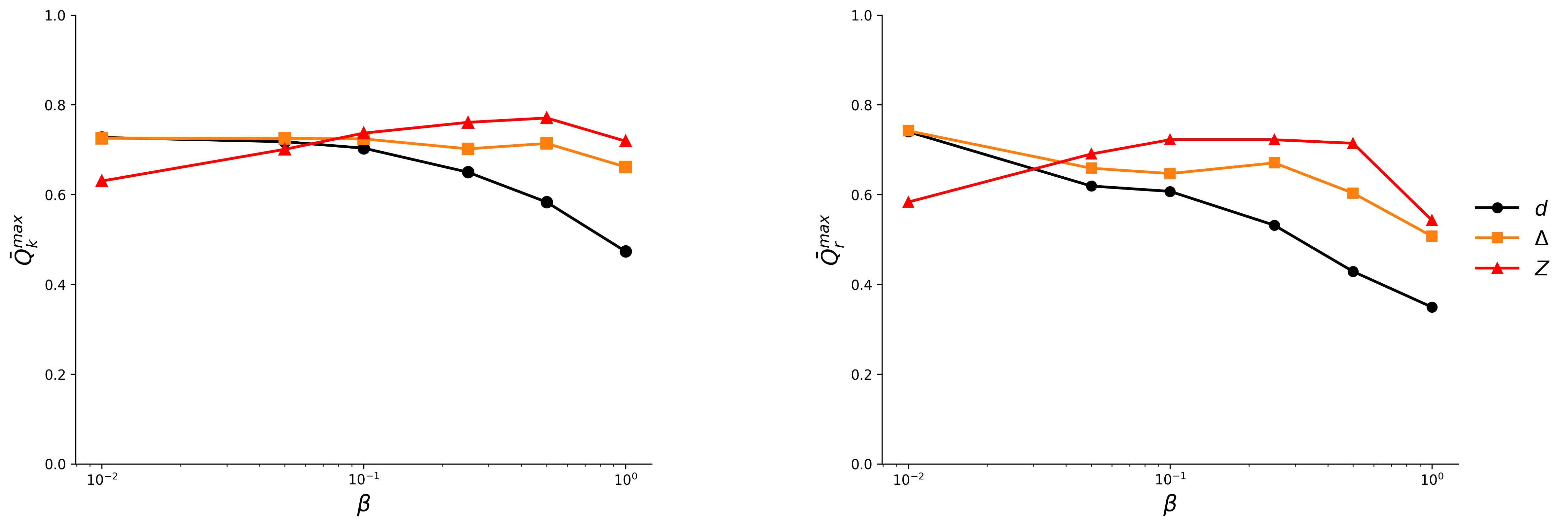}}
    \label{fig:subfig2}
  \end{minipage}
  \caption{The (best) prediction quality $\bar{Q}_k^{max}$ and $\bar{Q}_r^{max}$ of weighted degree mass ($d$), time-scaled temporal degree mass ($\Delta$), and time-scaled temporal reachability ($Z$), respectively, across various combinations of $\phi$ and $\beta$, in network HighScholl12.}
  \label{metrics performances of highschool12}
\end{figure*}

\begin{figure*}[h!]
  \centering
  \begin{minipage}[b]{1.0\textwidth}
    \subfigure[$\phi = 0.25$]{\includegraphics[width=\textwidth]{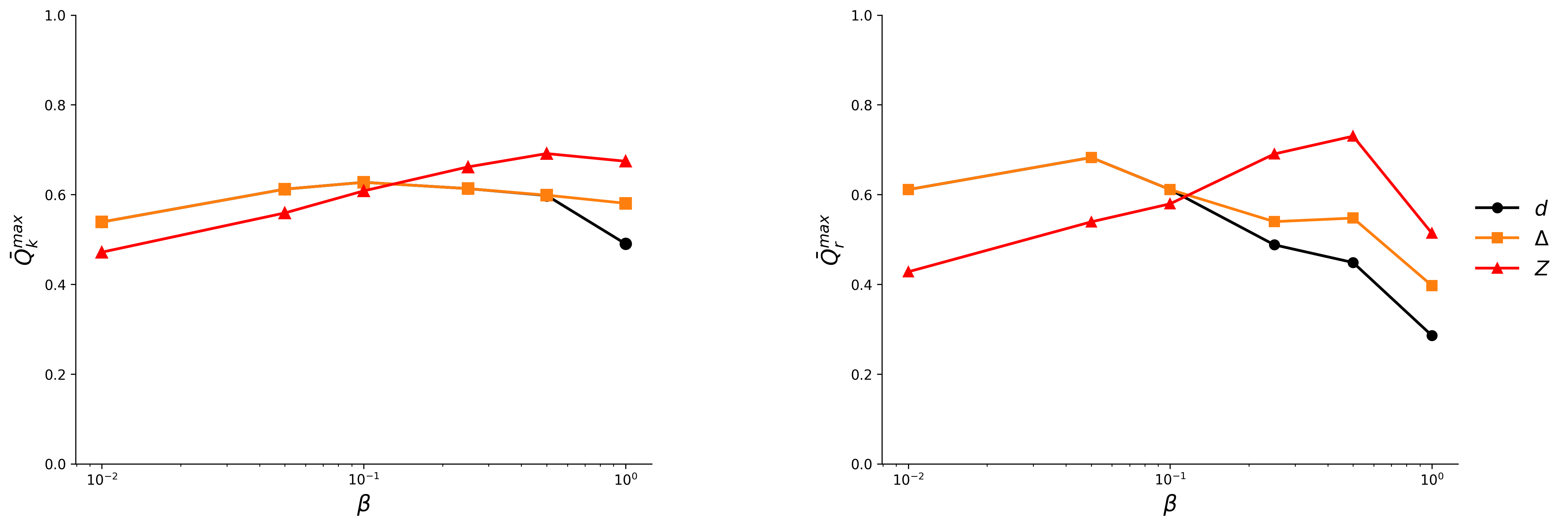}}
    \label{fig:subfig1}
  \end{minipage}
  \vspace{0.5cm}  % Optional: adds vertical space between the two images
  \begin{minipage}[b]{1.0\textwidth}
    \subfigure[$\phi = 0.5$]{\includegraphics[width=\textwidth]{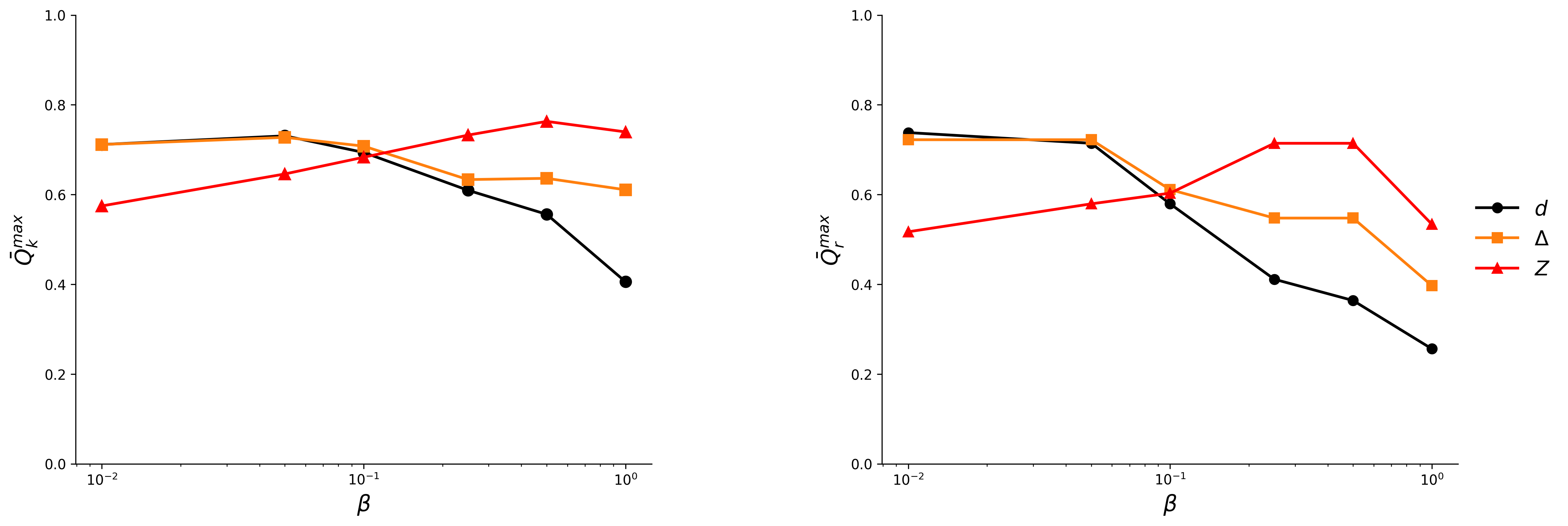}}
    \label{fig:subfig2}
  \end{minipage}
  \caption{The (best) prediction quality $\bar{Q}_k^{max}$ and $\bar{Q}_r^{max}$ of weighted degree mass ($d$), time-scaled temporal degree mass ($\Delta$), and time-scaled temporal reachability ($Z$), respectively, across various combinations of $\phi$ and $\beta$, in network Workplace13.}
  \label{metrics performances of work1}
\end{figure*}

\begin{figure*}[h!]
  \centering
  \begin{minipage}[b]{1.0\textwidth}
    \subfigure[$\phi = 0.25$]{\includegraphics[width=\textwidth]{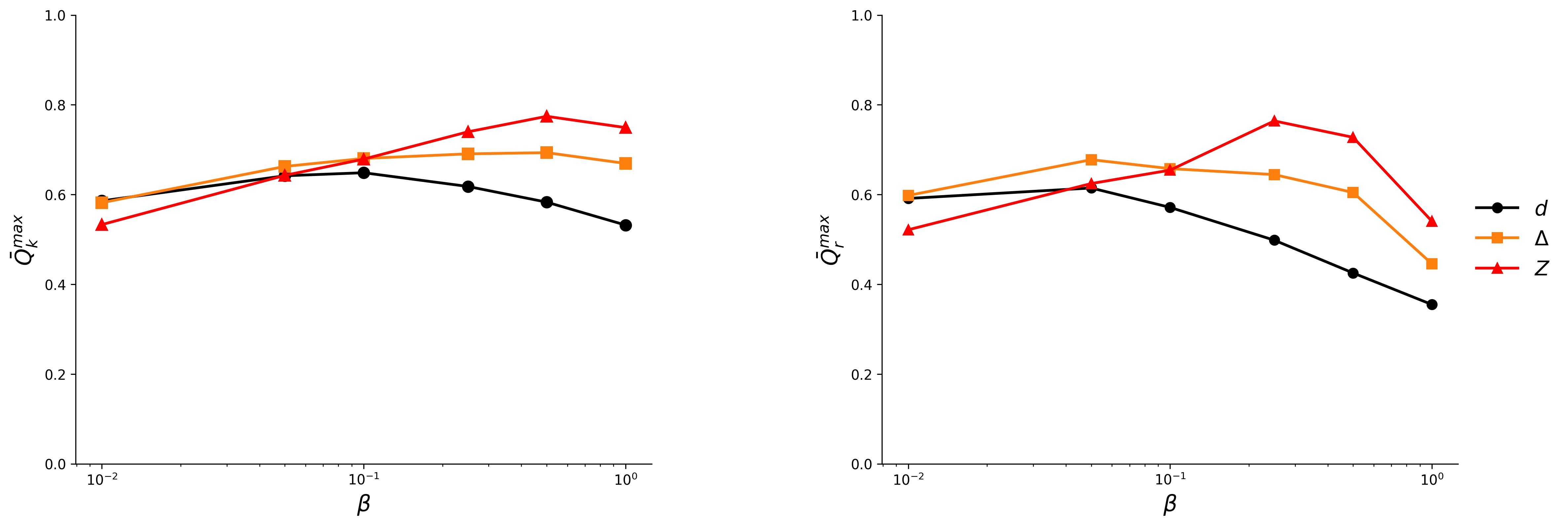}}
    \label{fig:subfig1}
  \end{minipage}
  \vspace{0.5cm}  % Optional: adds vertical space between the two images
  \begin{minipage}[b]{1.0\textwidth}
    \subfigure[$\phi = 0.5$]{\includegraphics[width=\textwidth]{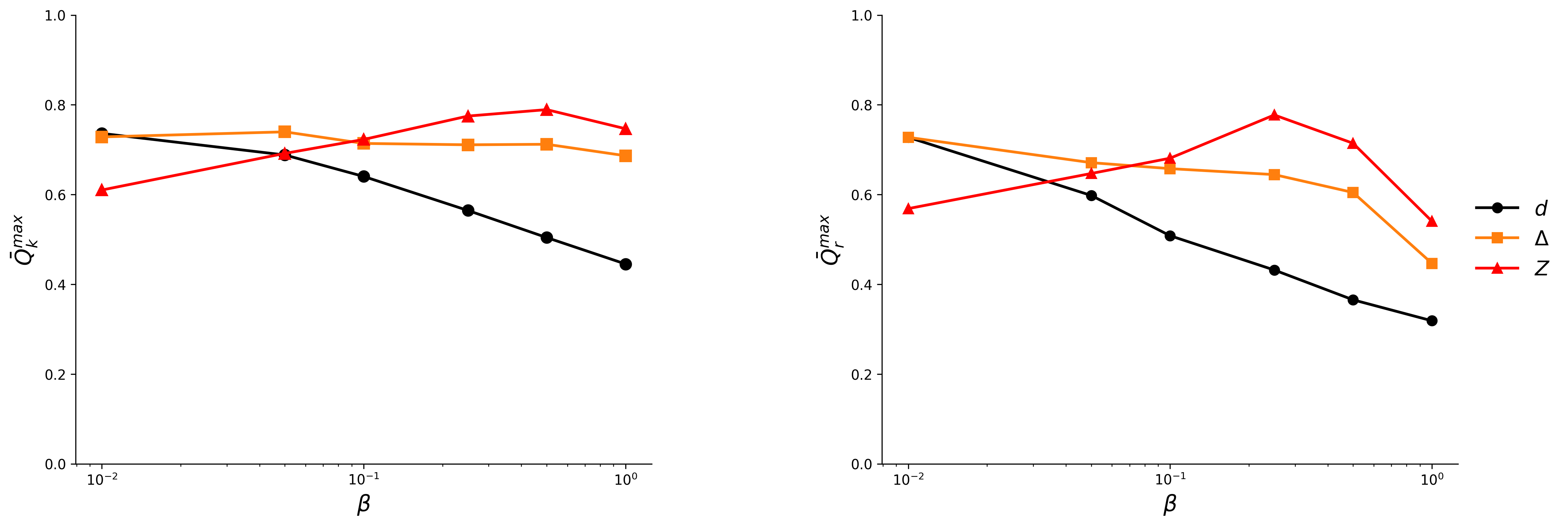}}
    \label{fig:subfig2}
  \end{minipage}
  \caption{The (best) prediction quality $\bar{Q}_k^{max}$ and $\bar{Q}_r^{max}$ of weighted degree mass ($d$), time-scaled temporal degree mass ($\Delta$), and time-scaled temporal reachability ($Z$), respectively, across various combinations of $\phi$ and $\beta$, in network Workplace15.}
  \label{metrics performances of work2}
\end{figure*}

\begin{figure*}[h!]
  \centering
  \begin{minipage}[b]{1.0\textwidth}
    \subfigure[$\phi = 0.25$]{\includegraphics[width=\textwidth]{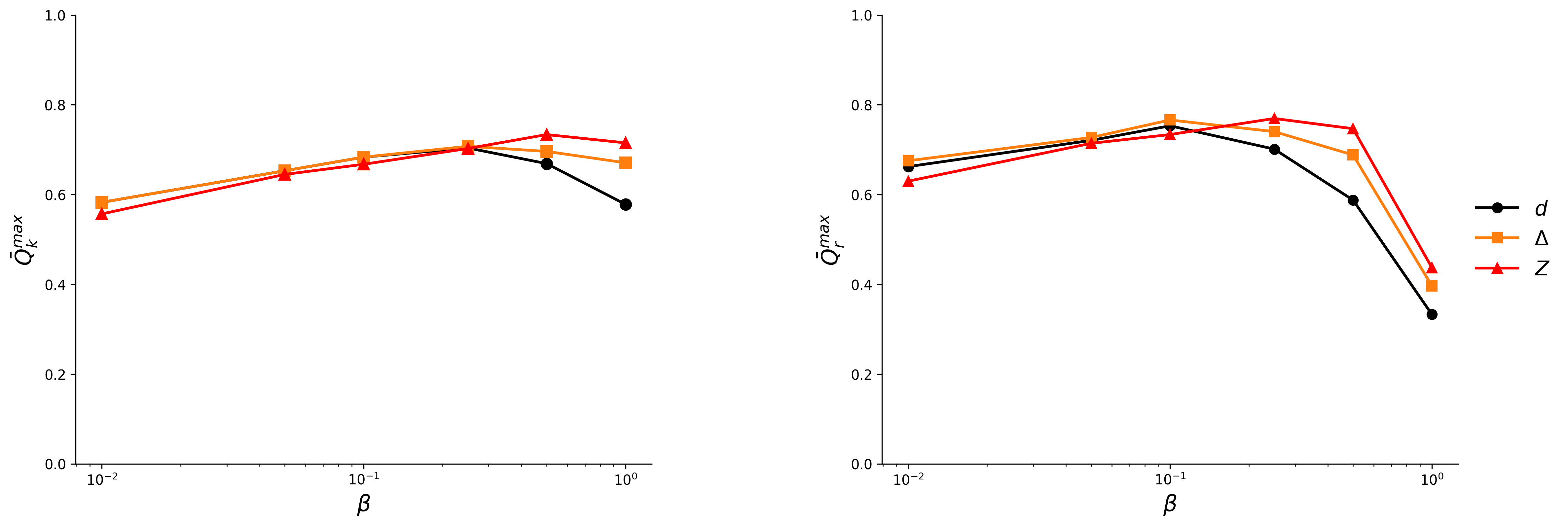}}
    \label{fig:subfig1}
  \end{minipage}
  \vspace{0.5cm}  % Optional: adds vertical space between the two images
  \begin{minipage}[b]{1.0\textwidth}
    \subfigure[$\phi = 0.5$]{\includegraphics[width=\textwidth]{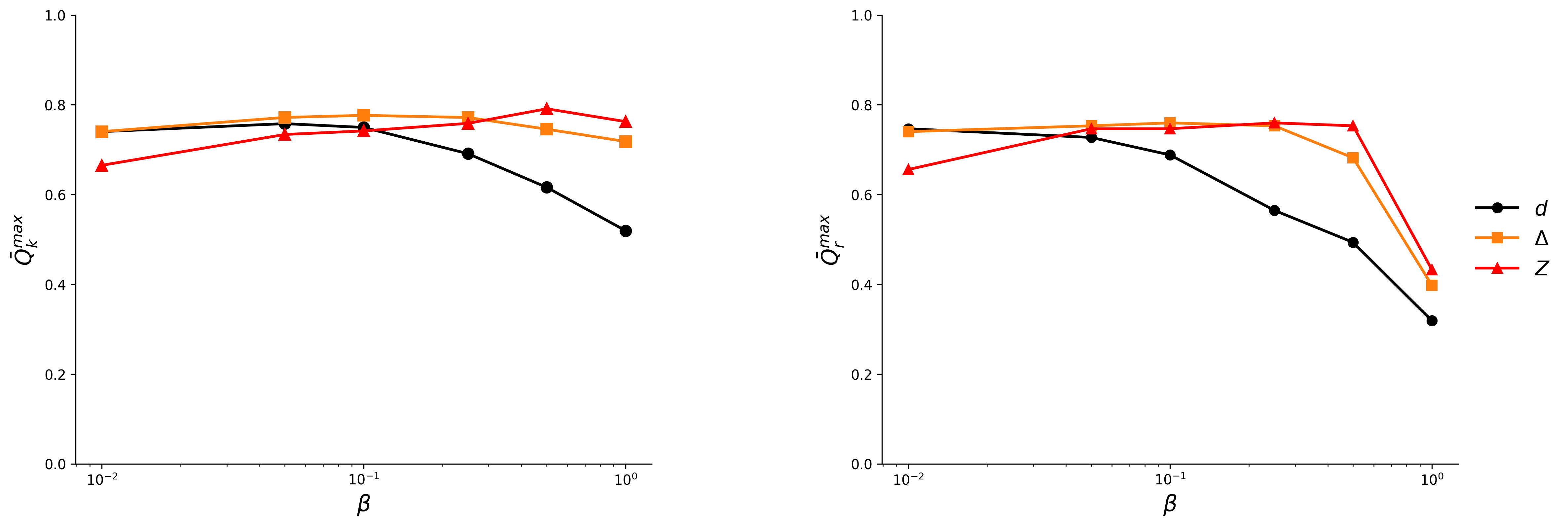}}
    \label{fig:subfig2}
  \end{minipage}
  \caption{The (best) prediction quality $\bar{Q}_k^{max}$ and $\bar{Q}_r^{max}$ of weighted degree mass ($d$), time-scaled temporal degree mass ($\Delta$), and time-scaled temporal reachability ($Z$), respectively, across various combinations of $\phi$ and $\beta$, in network Hyper-text.}
  \label{metrics performances of Hyper-text}
\end{figure*}

\begin{figure*}[h!]
  \centering
  \begin{minipage}[b]{1.0\textwidth}
    \subfigure[$\phi = 0.25$]{\includegraphics[width=\textwidth]{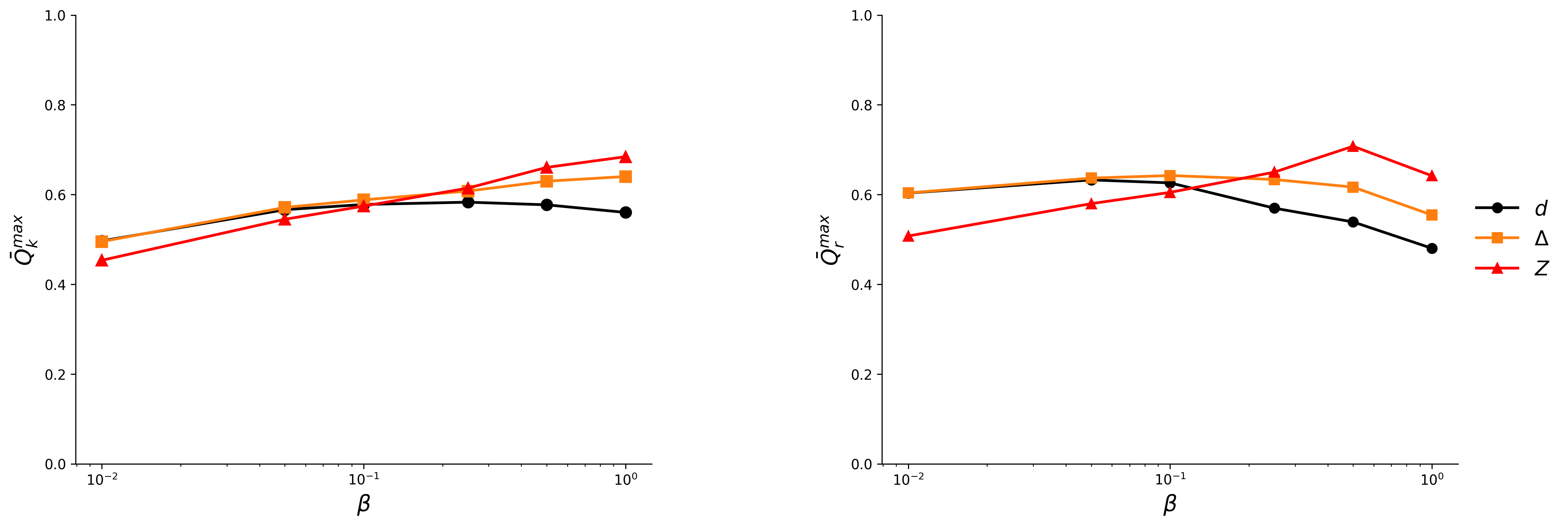}}
    \label{fig:subfig1}
  \end{minipage}
  \vspace{0.5cm}  % Optional: adds vertical space between the two images
  \begin{minipage}[b]{1.0\textwidth}
    \subfigure[$\phi = 0.5$]{\includegraphics[width=\textwidth]{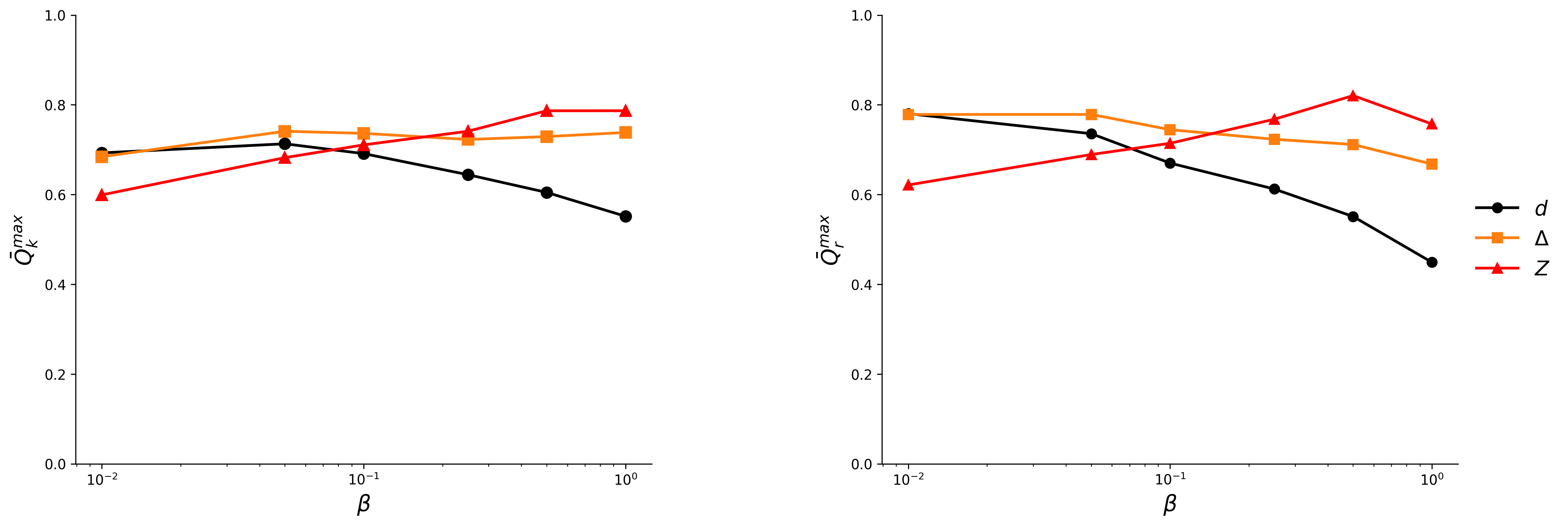}}
    \label{fig:subfig2}
  \end{minipage}
  \caption{The (best) prediction quality $\bar{Q}_k^{max}$ and $\bar{Q}_r^{max}$ of weighted degree mass ($d$), time-scaled temporal degree mass ($\Delta$), and time-scaled temporal reachability ($Z$), respectively, across various combinations of $\phi$ and $\beta$, in network SFHH.}
  \label{metrics performances of SFHH}
\end{figure*}

\begin{figure*}[h!]
  \centering
  \begin{minipage}[b]{1.0\textwidth}
    \subfigure[$\phi = 0.25$]{\includegraphics[width=\textwidth]{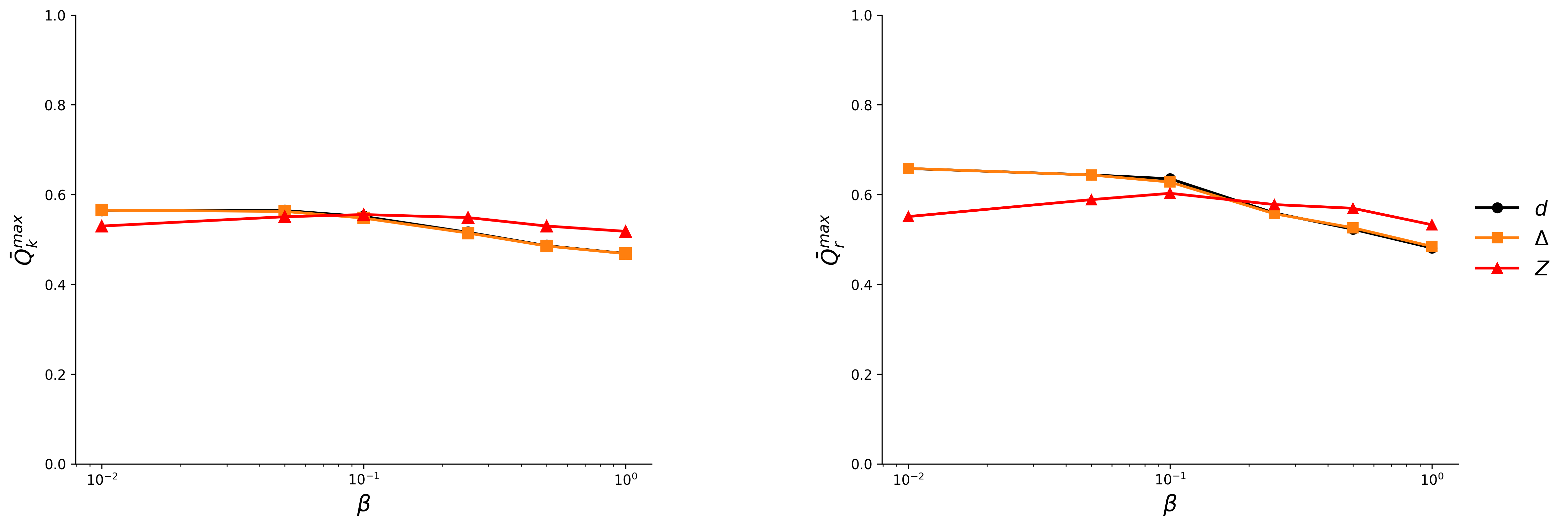}}
    \label{fig:subfig1}
  \end{minipage}
  \vspace{0.5cm}  % Optional: adds vertical space between the two images
  \begin{minipage}[b]{1.0\textwidth}
    \subfigure[$\phi = 0.5$]{\includegraphics[width=\textwidth]{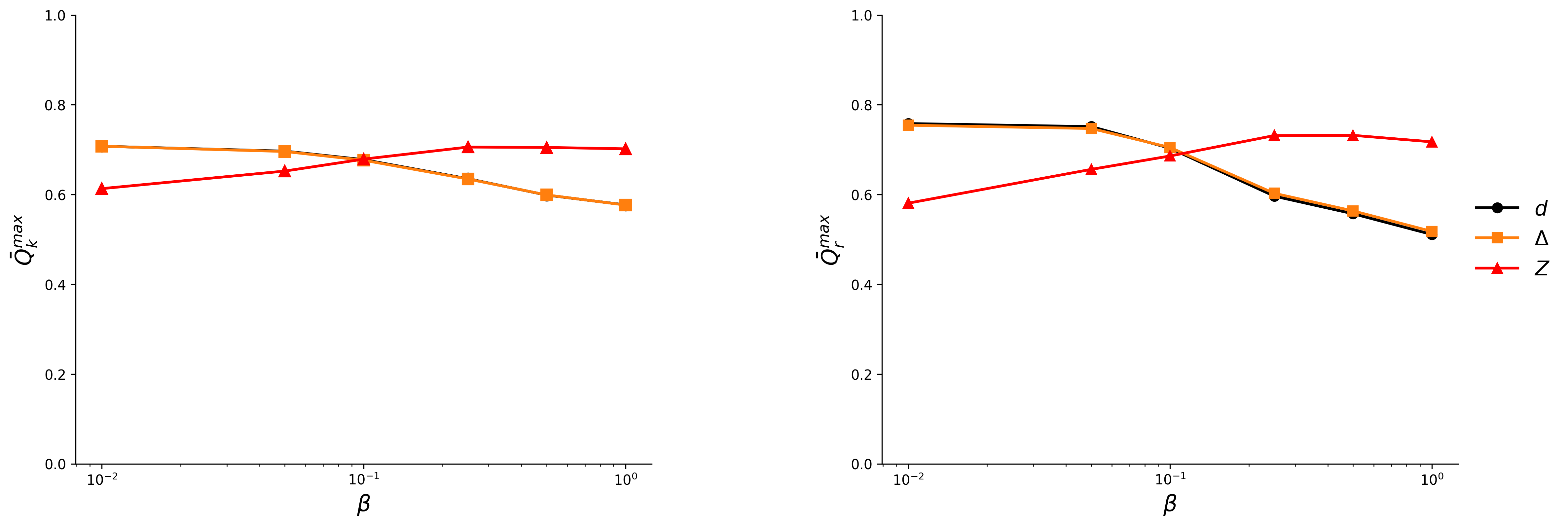}}
    \label{fig:subfig2}
  \end{minipage}
  \caption{The (best) prediction quality $\bar{Q}_k^{max}$ and $\bar{Q}_r^{max}$ of weighted degree mass ($d$), time-scaled temporal degree mass ($\Delta$), and time-scaled temporal reachability ($Z$), respectively, across various combinations of $\phi$ and $\beta$, in network Sms.}
  \label{metrics performances of Sms}
\end{figure*}

\begin{figure*}[h!]
  \centering
  \begin{minipage}[b]{1.0\textwidth}
    \subfigure[$\phi = 0.25$]{\includegraphics[width=\textwidth]{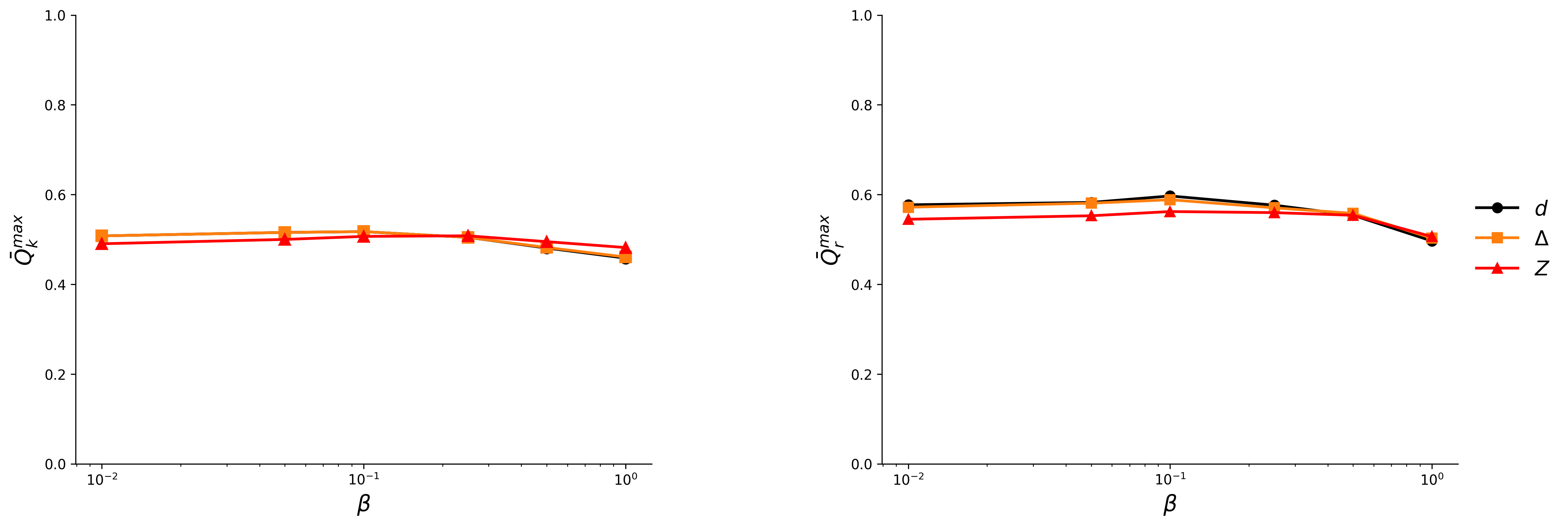}}
    \label{fig:subfig1}
  \end{minipage}
  \vspace{0.5cm}  % Optional: adds vertical space between the two images
  \begin{minipage}[b]{1.0\textwidth}
    \subfigure[$\phi = 0.5$]{\includegraphics[width=\textwidth]{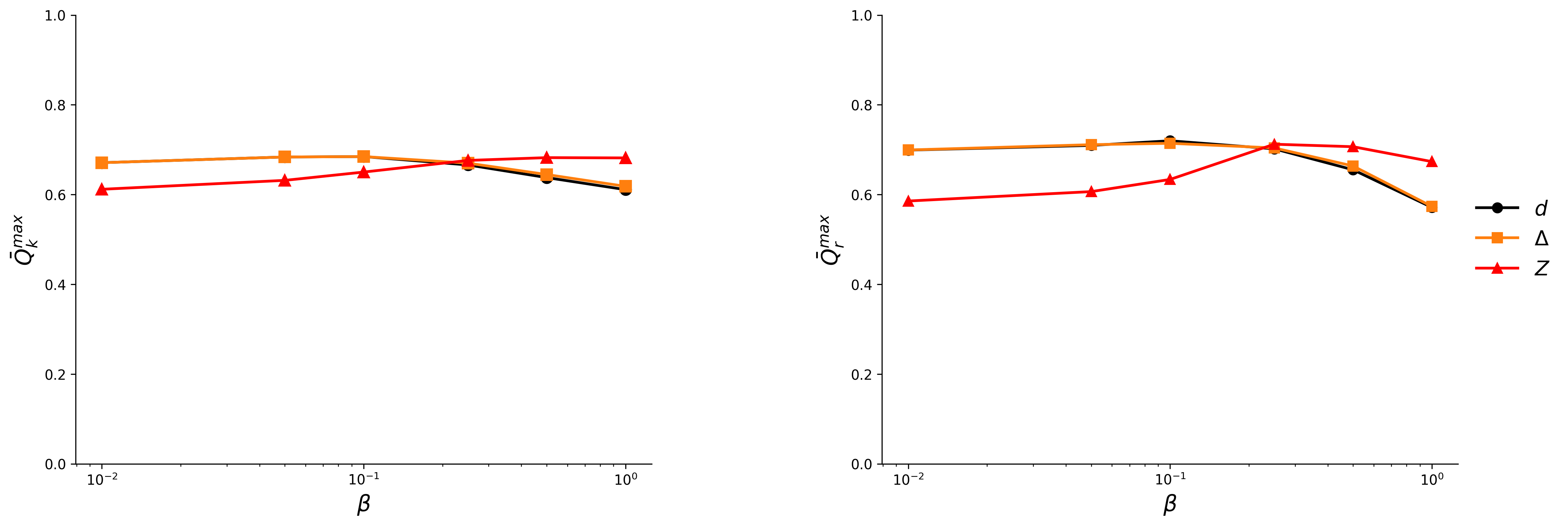}}
    \label{fig:subfig2}
  \end{minipage}
  \caption{The (best) prediction quality $\bar{Q}_k^{max}$ and $\bar{Q}_r^{max}$ of weighted degree mass ($d$), time-scaled temporal degree mass ($\Delta$), and time-scaled temporal reachability ($Z$), respectively, across various combinations of $\phi$ and $\beta$, in network Calls.}
  \label{metrics performances of Calls}
\end{figure*}

\begin{figure*}[h!]
  \centering
  \begin{minipage}[b]{1.0\textwidth}
    \subfigure[$\phi = 0.25$]{\includegraphics[width=\textwidth]{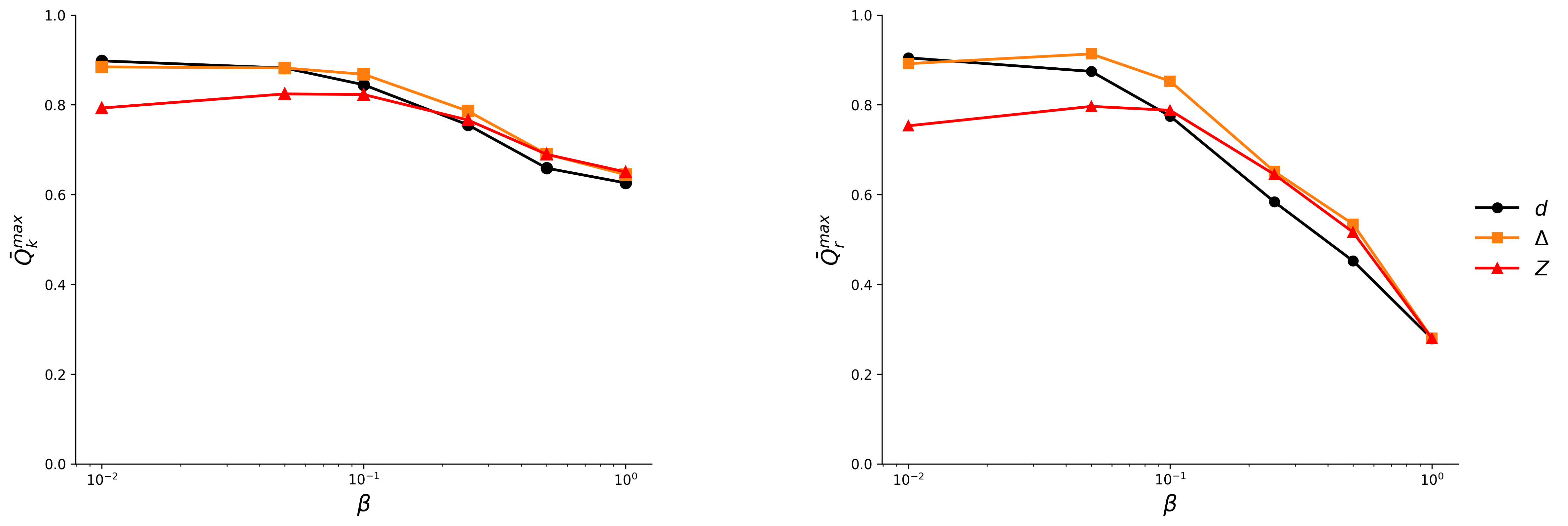}}
    \label{fig:subfig1}
  \end{minipage}
  \vspace{0.5cm}  % Optional: adds vertical space between the two images
  \begin{minipage}[b]{1.0\textwidth}
    \subfigure[$\phi = 0.5$]{\includegraphics[width=\textwidth]{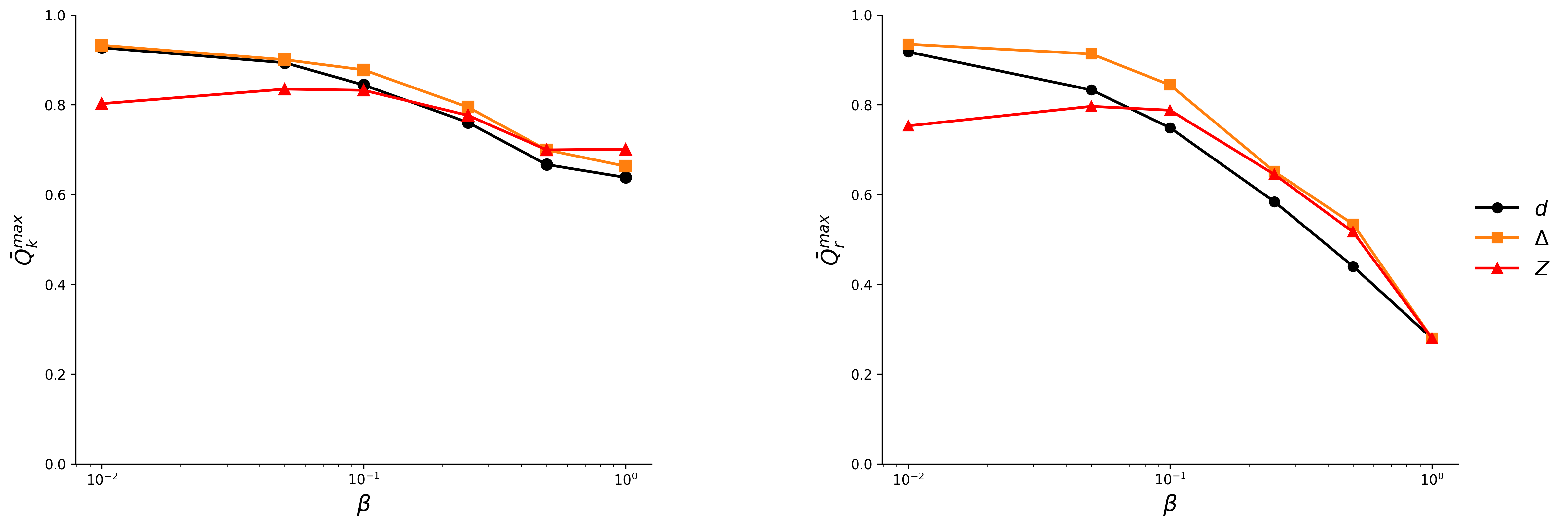}}
    \label{fig:subfig2}
  \end{minipage}
  \caption{The (best) prediction quality $\bar{Q}_k^{max}$ and $\bar{Q}_r^{max}$ of weighted degree mass ($d$), time-scaled temporal degree mass ($\Delta$), and time-scaled temporal reachability ($Z$), respectively, across various combinations of $\phi$ and $\beta$, in network Manufacturing Emails.}
  \label{metrics performances of Manufacturing Emails}
\end{figure*}

The appendix presents the prediction quality of three proposed centrality metrics in eight real-world networks, excluding HighSchool11 (see figures \ref{metrics performances of highschool12}-\ref{metrics performances of Manufacturing Emails}). Consistent with the observations in Section \ref{Prediction quality evaluation on real-world networks}, we observe the same in these networks as in HighSchool11: the time-scaled temporal reachability $Z$ generally achieves the best performance when $\beta$ is not small; However, for small $\beta$, $Z$ performs the worst whereas the other two metrics exhibit comparable performance. 

\begin{figure*}[htbp]
  \centering
  \includegraphics[width=\textwidth]{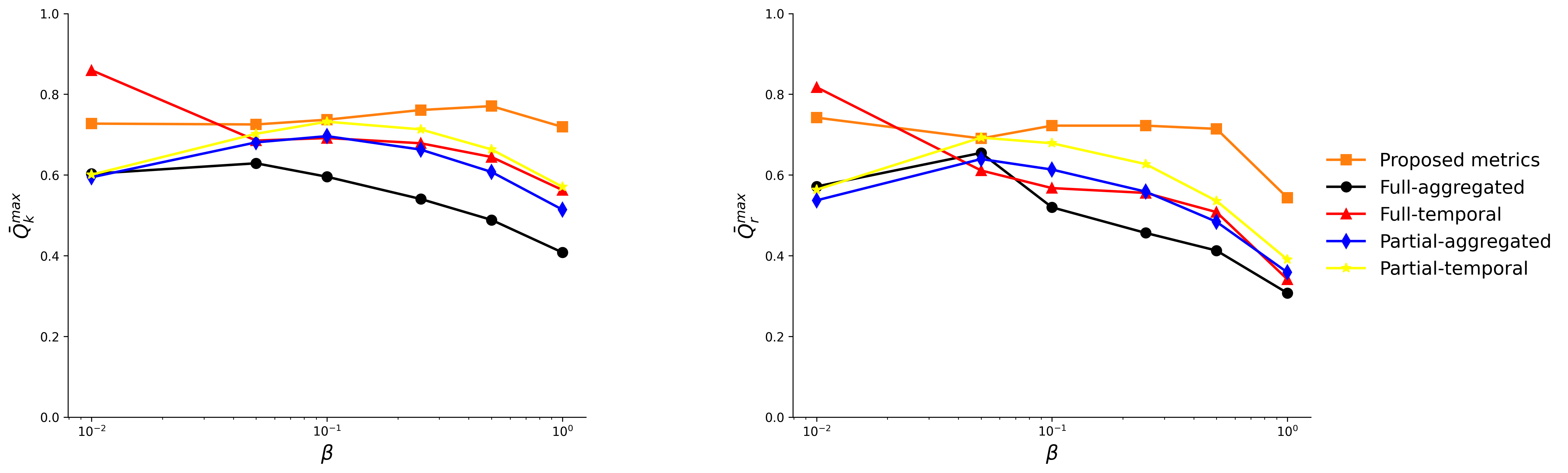}
  \vspace{2mm} % 增加间距，避免文字紧贴图像
  \text{(a) $\phi = 0.5$} % 单独添加 $\phi$ 参数下标
  \caption{
  The best prediction quality $\bar{Q}_k^{max}$ and $\bar{Q}_r^{max}$, respectively achieved by proposed centrality metrics derived from the partial temporal network $\mathcal{G}_{i}(\phi,m)$ (denoted by orange squares); full-aggregated: each static centrality metric derived from the unweighted aggregated network of the full temporal network $G$ (denoted by black dots); full-temporal: the average of each static centrality metric derived from all snapshots of $G$ or the temporal closeness centrality derived from $G$ (denoted by red triangles); partial-aggregated: each static centrality metric derived from the unweighted aggregated network of the partial temporal network $\mathcal{G}_{i}(\phi,m)$ (denoted by blue diamond); partial-temporal: the average of each static centrality metric derived from all snapshots of the partial temporal network $\mathcal{G}_{i}(\phi,m)$ or temporal closeness centrality derived from $\mathcal{G}_{i}(\phi,m)$ (denoted by yellow stars) when $\phi=0.5$ and $\beta$ varies, based on Highschool12.
  }
  \label{Prediction performance comparison between proposed local centrality metrics and classic global centrality metrics in Highschool12}
\end{figure*}

\begin{figure*}[htbp]
  \centering
  \includegraphics[width=\textwidth]{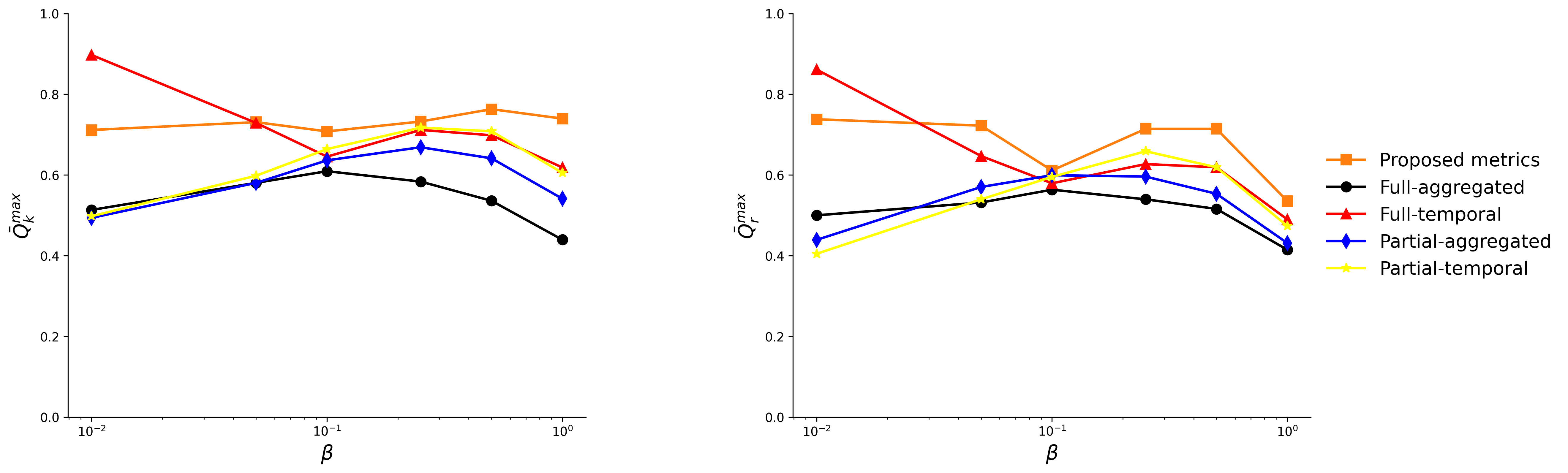}
  \vspace{2mm} % 增加间距，避免文字紧贴图像
  \text{(a) $\phi = 0.5$} % 单独添加 $\phi$ 参数下标
  \caption{
  The best prediction quality $\bar{Q}_k^{max}$ and $\bar{Q}_r^{max}$, respectively achieved by proposed centrality metrics derived from the partial temporal network $\mathcal{G}_{i}(\phi,m)$ (denoted by orange squares); full-aggregated: each static centrality metric derived from the unweighted aggregated network of the full temporal network $G$ (denoted by black dots); full-temporal: the average of each static centrality metric derived from all snapshots of $G$ or the temporal closeness centrality derived from $G$ (denoted by red triangles); partial-aggregated: each static centrality metric derived from the unweighted aggregated network of the partial temporal network $\mathcal{G}_{i}(\phi,m)$ (denoted by blue diamond); partial-temporal: the average of each static centrality metric derived from all snapshots of the partial temporal network $\mathcal{G}_{i}(\phi,m)$ or temporal closeness centrality derived from $\mathcal{G}_{i}(\phi,m)$ (denoted by yellow stars) when $\phi=0.5$ and $\beta$ varies, based on Workplace13.
  }
  \label{Prediction performance comparison between proposed local centrality metrics and classic global centrality metrics in work1}
\end{figure*}

\begin{figure*}[htbp]
  \centering
  \includegraphics[width=\textwidth]{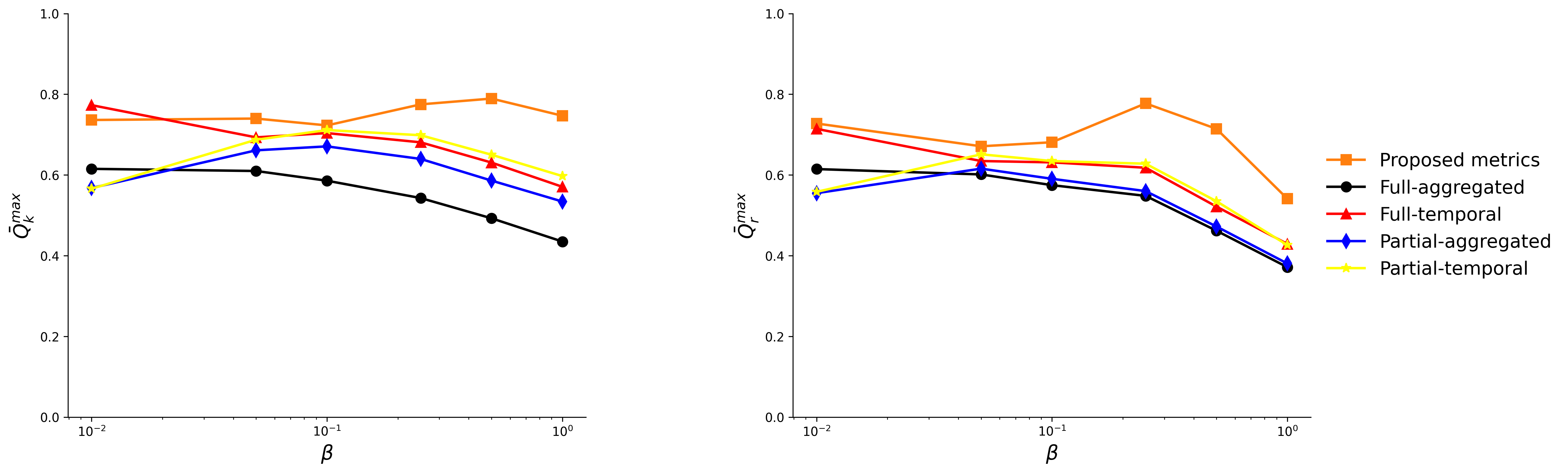}
  \vspace{2mm} % 增加间距，避免文字紧贴图像
  \text{(a) $\phi = 0.5$} % 单独添加 $\phi$ 参数下标
  \caption{
  The best prediction quality $\bar{Q}_k^{max}$ and $\bar{Q}_r^{max}$, respectively achieved by proposed centrality metrics derived from the partial temporal network $\mathcal{G}_{i}(\phi,m)$ (denoted by orange squares); full-aggregated: each static centrality metric derived from the unweighted aggregated network of the full temporal network $G$ (denoted by black dots); full-temporal: the average of each static centrality metric derived from all snapshots of $G$ or the temporal closeness centrality derived from $G$ (denoted by red triangles); partial-aggregated: each static centrality metric derived from the unweighted aggregated network of the partial temporal network $\mathcal{G}_{i}(\phi,m)$ (denoted by blue diamond); partial-temporal: the average of each static centrality metric derived from all snapshots of the partial temporal network $\mathcal{G}_{i}(\phi,m)$ or temporal closeness centrality derived from $\mathcal{G}_{i}(\phi,m)$ (denoted by yellow stars) when $\phi=0.5$ and $\beta$ varies, based on Workplace15.
  }
  \label{Prediction performance comparison between proposed local centrality metrics and classic global centrality metrics in work2}
\end{figure*}

\begin{figure*}[htbp]
  \centering
  \includegraphics[width=\textwidth]{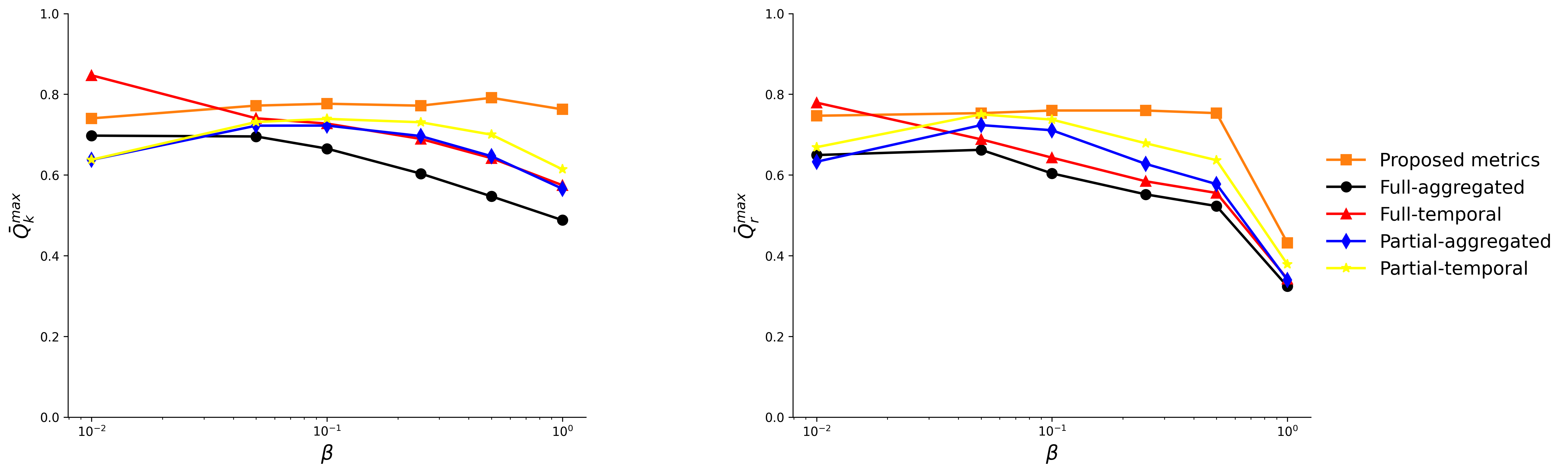}
  \vspace{2mm} % 增加间距，避免文字紧贴图像
  \text{(a) $\phi = 0.5$} % 单独添加 $\phi$ 参数下标
  \caption{
  The best prediction quality $\bar{Q}_k^{max}$ and $\bar{Q}_r^{max}$, respectively achieved by proposed centrality metrics derived from the partial temporal network $\mathcal{G}_{i}(\phi,m)$ (denoted by orange squares); full-aggregated: each static centrality metric derived from the unweighted aggregated network of the full temporal network $G$ (denoted by black dots); full-temporal: the average of each static centrality metric derived from all snapshots of $G$ or the temporal closeness centrality derived from $G$ (denoted by red triangles); partial-aggregated: each static centrality metric derived from the unweighted aggregated network of the partial temporal network $\mathcal{G}_{i}(\phi,m)$ (denoted by blue diamond); partial-temporal: the average of each static centrality metric derived from all snapshots of the partial temporal network $\mathcal{G}_{i}(\phi,m)$ or temporal closeness centrality derived from $\mathcal{G}_{i}(\phi,m)$ (denoted by yellow stars) when $\phi=0.5$ and $\beta$ varies, based on Hyper-text.
  }
  \label{Prediction performance comparison between proposed local centrality metrics and classic global centrality metrics in Hyper-text}
\end{figure*}

\begin{figure*}[htbp]
  \centering
  \includegraphics[width=\textwidth]{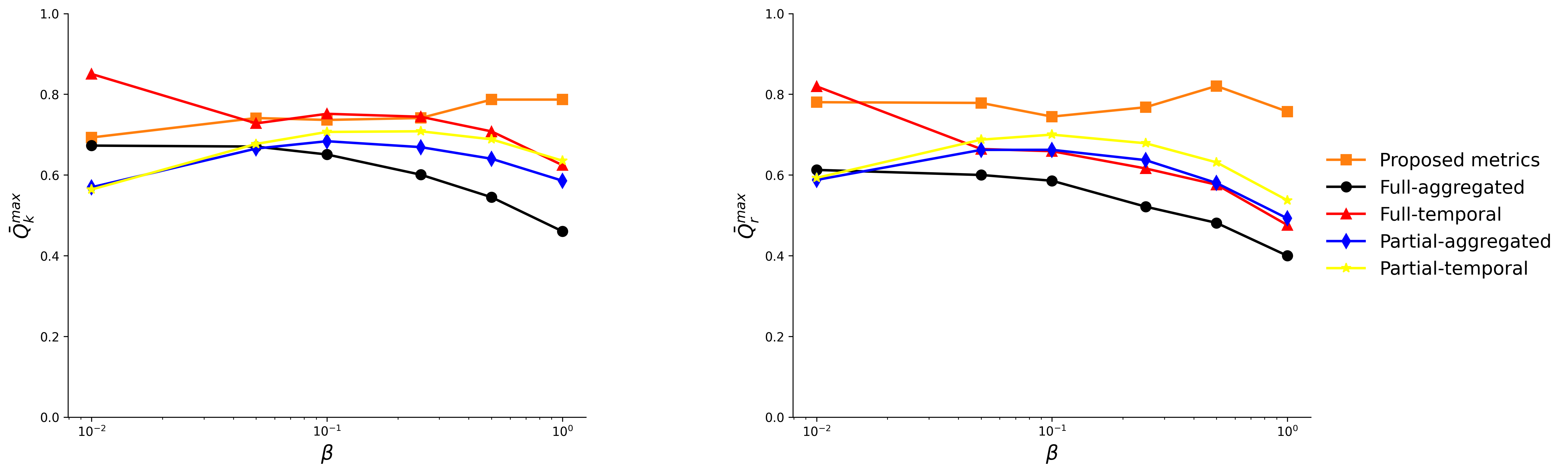}
  \vspace{2mm} % 增加间距，避免文字紧贴图像
  \text{(a) $\phi = 0.5$} % 单独添加 $\phi$ 参数下标
  \caption{
  The best prediction quality $\bar{Q}_k^{max}$ and $\bar{Q}_r^{max}$, respectively achieved by proposed centrality metrics derived from the partial temporal network $\mathcal{G}_{i}(\phi,m)$ (denoted by orange squares); full-aggregated: each static centrality metric derived from the unweighted aggregated network of the full temporal network $G$ (denoted by black dots); full-temporal: the average of each static centrality metric derived from all snapshots of $G$ or the temporal closeness centrality derived from $G$ (denoted by red triangles); partial-aggregated: each static centrality metric derived from the unweighted aggregated network of the partial temporal network $\mathcal{G}_{i}(\phi,m)$ (denoted by blue diamond); partial-temporal: the average of each static centrality metric derived from all snapshots of the partial temporal network $\mathcal{G}_{i}(\phi,m)$ or temporal closeness centrality derived from $\mathcal{G}_{i}(\phi,m)$ (denoted by yellow stars) when $\phi=0.5$ and $\beta$ varies, based on SFHH.
  }
  \label{Prediction performance comparison between proposed local centrality metrics and classic global centrality metrics in SFHH}
\end{figure*}

\begin{figure*}[htbp]
  \centering
  \includegraphics[width=\textwidth]{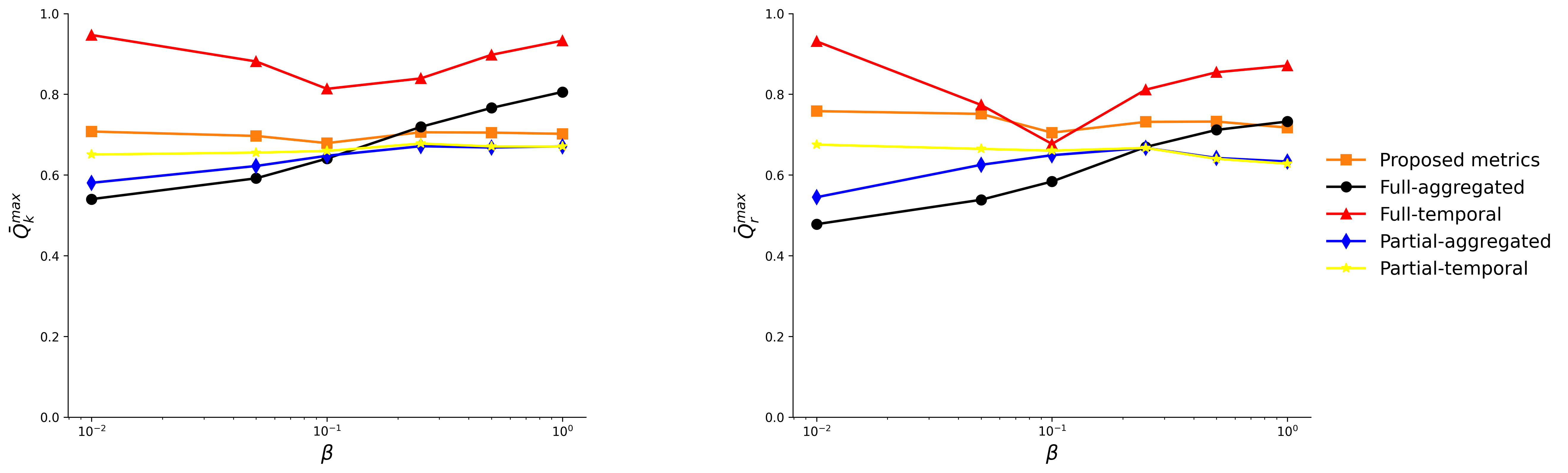}
  \vspace{2mm} % 增加间距，避免文字紧贴图像
  \text{(a) $\phi = 0.5$} % 单独添加 $\phi$ 参数下标
  \caption{
 The best prediction quality $\bar{Q}_k^{max}$ and $\bar{Q}_r^{max}$, respectively achieved by proposed centrality metrics derived from the partial temporal network $\mathcal{G}_{i}(\phi,m)$ (denoted by orange squares); full-aggregated: each static centrality metric derived from the unweighted aggregated network of the full temporal network $G$ (denoted by black dots); full-temporal: the average of each static centrality metric derived from all snapshots of $G$ or the temporal closeness centrality derived from $G$ (denoted by red triangles); partial-aggregated: each static centrality metric derived from the unweighted aggregated network of the partial temporal network $\mathcal{G}_{i}(\phi,m)$ (denoted by blue diamond); partial-temporal: the average of each static centrality metric derived from all snapshots of the partial temporal network $\mathcal{G}_{i}(\phi,m)$ or temporal closeness centrality derived from $\mathcal{G}_{i}(\phi,m)$ (denoted by yellow stars) when $\phi=0.5$ and $\beta$ varies, based on Sms.
  }
  \label{Prediction performance comparison between proposed local centrality metrics and classic global centrality metrics in Sms}
\end{figure*}

\begin{figure*}[htbp]
  \centering
  \includegraphics[width=\textwidth]{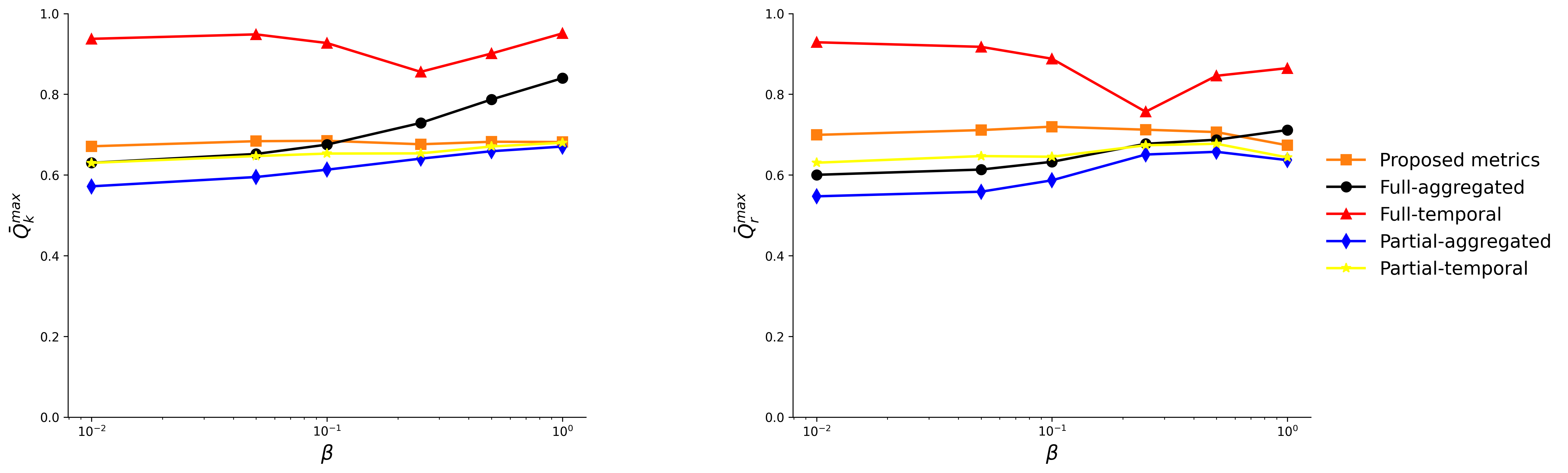}
  \vspace{2mm} % 增加间距，避免文字紧贴图像
  \text{(a) $\phi = 0.5$} % 单独添加 $\phi$ 参数下标
  \caption{
  The best prediction quality $\bar{Q}_k^{max}$ and $\bar{Q}_r^{max}$, respectively achieved by proposed centrality metrics derived from the partial temporal network $\mathcal{G}_{i}(\phi,m)$ (denoted by orange squares); full-aggregated: each static centrality metric derived from the unweighted aggregated network of the full temporal network $G$ (denoted by black dots); full-temporal: the average of each static centrality metric derived from all snapshots of $G$ or the temporal closeness centrality derived from $G$ (denoted by red triangles); partial-aggregated: each static centrality metric derived from the unweighted aggregated network of the partial temporal network $\mathcal{G}_{i}(\phi,m)$ (denoted by blue diamond); partial-temporal: the average of each static centrality metric derived from all snapshots of the partial temporal network $\mathcal{G}_{i}(\phi,m)$ or temporal closeness centrality derived from $\mathcal{G}_{i}(\phi,m)$ (denoted by yellow stars) when $\phi=0.5$ and $\beta$ varies, based on Calls.
  }
  \label{Prediction performance comparison between proposed local centrality metrics and classic global centrality metrics in Calls}
\end{figure*}

\begin{figure*}[htbp]
  \centering
  \includegraphics[width=\textwidth]{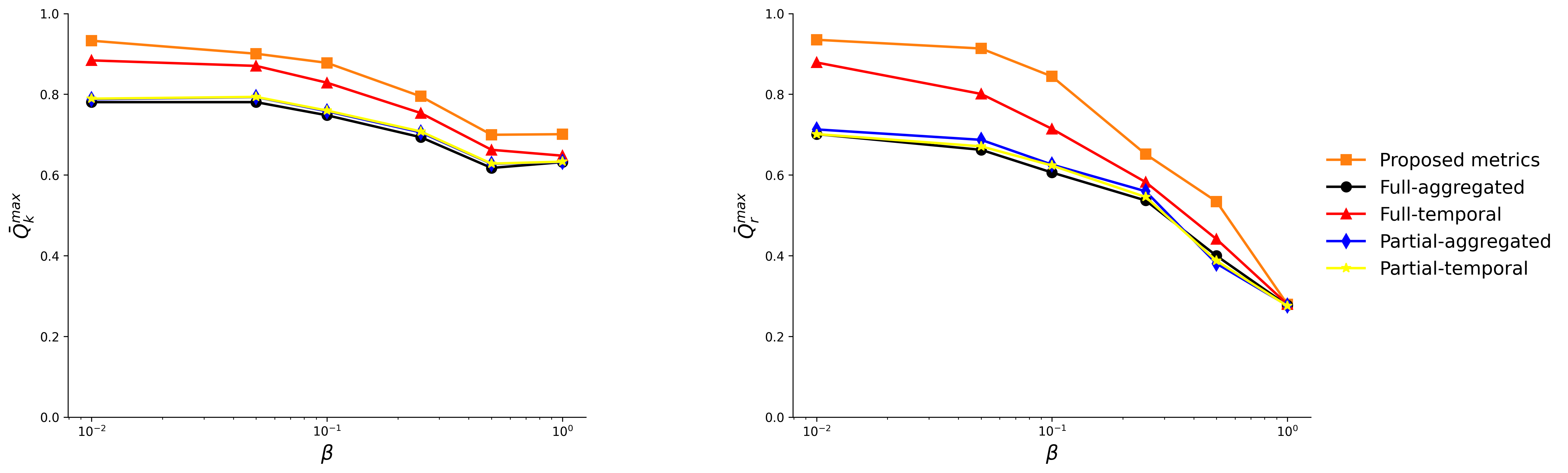}
  \vspace{2mm} % 增加间距，避免文字紧贴图像
  \text{(a) $\phi = 0.5$} % 单独添加 $\phi$ 参数下标
  \caption{
  The best prediction quality $\bar{Q}_k^{max}$ and $\bar{Q}_r^{max}$, respectively achieved by proposed centrality metrics derived from the partial temporal network $\mathcal{G}_{i}(\phi,m)$ (denoted by orange squares); full-aggregated: each static centrality metric derived from the unweighted aggregated network of the full temporal network $G$ (denoted by black dots); full-temporal: the average of each static centrality metric derived from all snapshots of $G$ or the temporal closeness centrality derived from $G$ (denoted by red triangles); partial-aggregated: each static centrality metric derived from the unweighted aggregated network of the partial temporal network $\mathcal{G}_{i}(\phi,m)$ (denoted by blue diamond); partial-temporal: the average of each static centrality metric derived from all snapshots of the partial temporal network $\mathcal{G}_{i}(\phi,m)$ or temporal closeness centrality derived from $\mathcal{G}_{i}(\phi,m)$ (denoted by yellow stars) when $\phi=0.5$ and $\beta$ varies, based on Manufacturing Emails.
  }
  \label{Prediction performance comparison between proposed local centrality metrics and classic global centrality metrics in Manufacturing Emails}
\end{figure*}

The comparison between the proposed centrality metrics and classic centrality metrics in the remaining eight real-world networks is shown in figures \ref{Prediction performance comparison between proposed local centrality metrics and classic global centrality metrics in Highschool12}-\ref{Prediction performance comparison between proposed local centrality metrics and classic global centrality metrics in Manufacturing Emails}. The finding that proposed centrality metrics mostly outperform classic centrality metrics in HighSchool11, discussed in Section \ref{Comparison with classic centrality metrics in prediction quality}, also holds for these eight real-world networks.

\clearpage
\bibliographystyle{plainnat}
\bibliography{main}  

\end{document}